\documentclass[onecolumn,showpacs]{revtex4}

\usepackage{graphicx}
\begin{document}
\title{The frequency spectrum of focused broadband pulses of electromagnetic
radiation generated by polarization currents with superluminally rotating distribution patterns}

\author{H. Ardavan$^1$, A. Ardavan$^2$ and J. Singleton$^3$}
\affiliation{$^1$Institute of Astronomy, University of Cambridge,
Madingley Road, Cambridge CB3 0HA, United Kingdom\\
$^2$Clarendon Laboratory, Department of Physics, University of Oxford,
Parks Road, Oxford OX1 3PU, United Kingdom\\
$^3$National High Magnetic Field Laboratory, TA-35, MS-E536,
Los Alamos National Laboratory, Los Alamos, NM87545, USA}
\begin{abstract}
We investigate the spectral features of the emission from a
superluminal polarization current whose distribution
pattern rotates with an angular frequency $\omega$
and oscillates with an incommensurate frequency $\Omega >\omega$.
This type of polarization current is found in recent practical
machines designed to investigate superluminal emission.
Although all of the processes involved are linear,
we find that the broadband emission
contains frequencies that are higher than $\Omega$ by a factor of the order of $(\Omega/\omega)^2$.
This generation of frequencies {\it not} required for the creation of the source
stems from mathematically rigorous consequences of the familiar classical
expression for the retarded potential.
The results suggest practical applications for superluminal polarization currents
as broad-band radiofrequency and infrared sources.
\end{abstract}
%\ocis{230.6080, 030.1670, 040.3060, 250.5530, 260.2110, 999.9999 Superluminal
%emission.} 

\maketitle

\section{Introduction}
Considerable recent interest has been generated by the design~\cite{AAS12,PhysWorld,NewSci,economist},
construction~\cite{johan} and testing~\cite{test} of novel
light sources which employ extended distributions of polarization currents moving faster
than $c$, the speed of light {\it in vacuo}.
Although Special Relativity does not allow a charged particle with finite inertial mass to move faster than $c$,
there is no such restriction on macroscopic polarization currents, because their superluminally moving distribution
patterns may be created by the coordinated motion of aggregates of subluminally moving particles~\cite{AAS1,hewish,hannay,reply,AAS2,AAS22,AAS23}.
Moreover, the distinction between massive and massless charges is not in any way reflected in Maxwell's equations,
so that the propagating distribution patterns of such polarization currents radiate as would
any other moving sources of the electromagnetic field~\cite{AAS1,AAS2,AAS22,AAS23}.

In view of this interest, the present paper explores the frequency
spectrum produced by such superluminal sources.
We base our analysis on the practical implementation of a superluminal
source (Appendix A,~\cite{AAS12,johan}), and consider the electromagnetic waves
emitted from a polarization current whose
distribution pattern {\it rotates} with an angular frequency $\omega$
and {\it oscillates} with an incommensurate frequency $\Omega>\omega$ (Table~1).
It is found that the broadband signals carried by such
waves contain frequencies that are by a factor of the order of $(\Omega/\omega)^2$
higher than the frequency $\Omega$; {\it i.e. the radiation
contains frequencies which are not required for the creation or the practical implementation of the source.}
This results from the cooperation of two effects,
both of which are mathematically rigorous consequences of the familiar classical
expression for the retarded potential.
\begin{enumerate}
\item
The retarded time is a multi-valued function
of the observation time in the superluminal regime, so that the interval of
retarded time during which a particular set of wave fronts is emitted by a
source point can be significantly longer than the interval of observation time during
which the same set of wave fronts is received at the observation point.
\item
The centripetal acceleration enriches the spectral content
of a rotating volume source, for which $\Omega/\omega$ is different from an integer,
by effectively endowing the distribution of its density with space-time discontinuities.
\end{enumerate}
These two effects make it possible, for instance, to generate a broadband pulse of radiation whose spectrum
has a peak in the terahertz band by means of a device whose construction and operation
only entails oscillations at two radio frequencies:  at $\Omega/(2\pi)=302.5$~MHz and at a
multiple $m\omega$ (with, e.g., $m=72$) of $\omega/(2\pi)=5$~MHz (see Table 1 and Appendix A).

\begin{table}[tbp] \centering
\begin{tabular}{|l|l|}
\hline
 Symbol  & Definition \\ \hline
 $\omega$ &  Angular rotation frequency of the distibution pattern of the source\\ \hline
 $\Omega$ &  The angular frequency with which the source oscillates (in addition to moving)\\ \hline
 $f$ &  Frequency of the radiation generated by the source \\ \hline
 $n=2\pi f/\omega$ &   The harmonic number associated with the radiation frequency\\ \hline
 $m$ &  The number of cycles of the sinusoidal wave train representing the azimuthal\\
 ~ & dependence of the rotating source distribution [see Eq.\ (7)] around the \\
 ~& circumference of a circle centred on, and normal to, the rotation axis \\ \hline
 $\vert\Omega\pm m\omega\vert$ &  The two frequencies to which the spectrum of the spherically decaying component \\
 ~ & of the radiation is limited when $\Omega/\omega$ is an integer.  (The non-spherically decaying\\
 ~ & component of the radiation is only emitted at these two frequencies, irrespective\\
 ~ & of whether $\Omega/\omega$ is an integer or not.)\\ \hline
\end{tabular}
\caption{Definition of the various frequencies and numbers
used to describe the source and the emitted radiation}
\label{table1}
\end{table}

This paper is organised as follows.
Section~\ref{IIA} outlines the novel features of the radiation from oscillating, accelerated superluminal polarization
currents, contrasting this type of emission with those of fast travelling wave antennas,
and \v Cerenkov and synchrotron radiations.
Although considerable mathematical rigour is required to derive the frequency spectrum of
such superluminal sources (this stems from the necessary three-dimensionality
of superluminal sources;
superluminal sources cannot be point-like~\cite{AAS1,AAS2,AAS22,AAS23}), some of the essential features
can be demonstrated in the case of a {\it localized} source, a source whose dimensions are appreciably smaller than
the radiation wavelength; this is carried out in Section~\ref{IIB} as an aid to understanding the more complex analysis that follows.
Section~\ref{IIC} introduces those properties of a rotating {\it volume}
source that contribute to to the distinctive spectral features of the emitted radiation.
The algebraic description of the polarization current (its time dependence and orientation) used in the
subsequent analysis is given in Section~\ref{IID}; all of the later results are presented in terms of the
parameters defined here.  Finally, Section~\ref{IIE} emphasizes the important contribution
that the {\it centripetal acceleration} makes to the broadband nature of the emission.
The detailed mathematical formulation of the problem of an extended superluminal source is set out in Section~\ref{III}.
The special case of those volume elements which contribute {\it coherently}
to the field at the observer is discussed in Section~\ref{IV}, highlighting the importance of focal regions in the space of observation points (Section~\ref{IVA}) and the inadequacies of conventional far-field approximations
in evaluating the electromagnetic fields (Section~\ref{IVB}).
The radiation field in the plane of the source's orbit is treated in Section~\ref{V},
which contains graphical representations of the spectral distribution of the
emitted power (Figs.\ 8 and 9);
the predicted frequency spectra show broadband, high-frequency emission from
a source whose implementation entails only two basic frequencies.
A similar treatment for the radiation field outside the plane of the source's
orbit is given in Section~\ref{VI}, which also describes the polarization of the
emitted radiation (Table~2).
A detailed comparison with synchrotron and dipole radiations using an analogous mathematical
description is given in Section~\ref{VII},
and the efficiency of the radiative process is estimated in Section~\ref{VIII}.
Conclusions and a summary are given in Section~\ref{IX}.
\section{Distinctive features of the emission from an extended, rotating, oscillating superluminal polarization current}
\label{II}
\subsection{Comparison with fast travelling wave antennas, and \v Cerenkov and synchrotron radiations}
\label{IIA}
Superluminally moving surface charges are already encountered in fast travelling wave antennas~\cite{AAS3} and certain types of leaky waveguides~\cite{AAS32}.  The wave fronts emanating from such uniformly moving
superluminal sources possess an envelope on which the vacuum version of the \v Cerenkov effect can be observed.
However, the source considered in the present paper has the time
dependence of a travelling wave with both an accelerated superluminal motion and an oscillating amplitude;
neither of these features are present in extant antennas.

The superluminal sources considered in the present paper have distinctive features of
both the sources of \v Cerenkov and synchrotron radiations:
not only do they move with a speed that exceeds the propagation speed of the
waves they generate (like sources of \v Cerenkov radiation)
but also their motion is centripetally accelerated (like that of sources of synchrotron radiation).
However, in contrast to sources of these familiar types of emission, they are extended (rather than point-like)
and they fluctuate in their strength (rather than being time-independent in their own rest frames).

Acceleration leads to the formation of a cusp in the envelope of the wave fronts that emanate from
each volume element of the superluminal source~\cite{AAS4}, a curve along which two sheets of
the envelope meet tangentially (Fig.\ 1). (By contrast, the conical envelope of wave fronts in
the \v Cerenkov emission has no cusp.)
At any given observation time, the radiated field entails a set of such envelopes
(each associated with the wave fronts emanating from a specific member of a
corresponding set of source elements) whose cusps pass through the observation point.
These caustics arise from those volume elements of the source which approach the observer, along
the radiation direction, with the speed of light and zero acceleration at the retarded time (Fig.\ 2).
The contribution from the filamentary locus of such source elements toward the value
of the field at the observation point has certain unexpected properties.
Its intensity, for instance, does not diminish with the distance $R_P$ from the source like ${R_P}^{-2}$,
as in the case of a spherically spreading wave, but more slowly:
like ${R_P}^{-\kappa}$ with $0<\kappa<2$ (see~\cite{AAS1}).

\begin{figure}[tbp]
   \centering
\includegraphics[height=7cm]{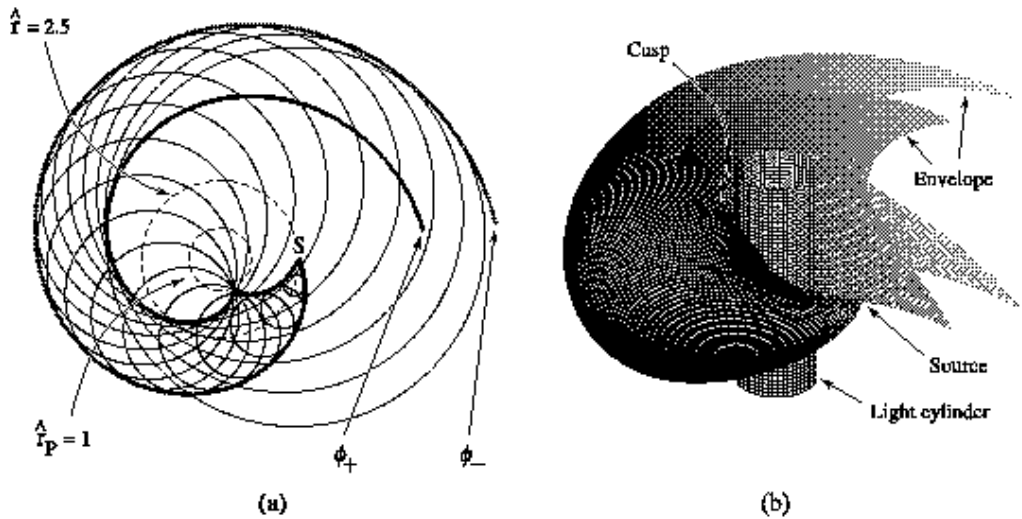}
\caption{(a) Envelope of the spherical wave fronts emanating from a source point $S$ which moves with a
constant angular velocity $\omega$ on a circle of radius $r=2.5 c/\omega$ (${\hat r}\equiv r\omega/c=2.5$).
The circles in broken lines designate the orbit of $S$ and the light cylinder $r_P=c/\omega$ (${\hat r}_P=1$).
The curves to which the emitted wave fronts are tangent are the cross sections of the two sheets $\phi_\pm$
of the envelope with the plane of source's orbit.  (b) Three-dimensional view of the light cylinder and the
envelope of wave fronts for the same source point $S$.  The tube-like surface constituting the
envelope is symmetric with respect to the plane of the orbit.  The cusp along which the two sheets of
this envelope meet touches, and is tangential to, the light cylinder at a point on the plane of the source's
orbit and spirals around the rotation axis out into the radiation zone (based on~\cite{AAS1}).}
\end{figure}

\begin{figure}[tbp]
   \centering
\includegraphics[height=6cm]{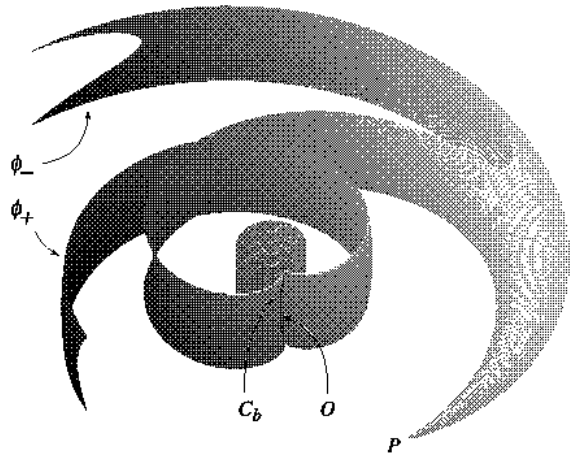}
\caption{The bifurcation surface (i.e.\ the locus of source points that approach
the observer along the radiation direction with the speed of light at the retarded time)
associated with the observation point $P$ for a clockwise source motion.
The cusp $C_b$, along which the two sheets of the bifurcation surface meet tangentially,
touches the light cylinder (${\hat r}=1$) at point $O$.
This cusp curve is the locus of source points which approach the observer not
only with the speed of light but also with zero acceleration along the radiation direction.
For an observation point in the radiation zone, the spiralling surface that issues from $P$ undergoes a
large number of turns, in which its two sheets intersect one another, before reaching the light cylinder.
Note that the bifurcation surface issues from the observation point $P$ and resides in the space $(r,{\hat\varphi},z)$ of source points, while the envelope of wave fronts issues from a source point and resides in the space $(r_P,{\hat\varphi}_P,z_P)$ of observation points; the similarity between these two surfaces reflects the reciprocity properties of the Green's function for the problem
(based on~\cite{AAS1}).}
\end{figure}

In the current paper, we shall restrict our analysis to the frequency content
of the spherically-decaying emission; the spectrum of the non-spherically-decaying
component of the emission is treated in a separate work~\cite{newpaper}.
As has already been mentioned, a rigorous analysis of this problem
is of considerable complexity (Sections~\ref{III}-\ref{VIII}) owing to the extended nature of superluminal
sources; we therefore first briefly discuss the case of a localized source,
a source whose dimensions are appreciably smaller than
the radiation wavelength. This simplified case defines the Green's function for the
more complex problem, and illustrates some properties of the emission which are also valid in the
more general case.
\subsection{Properties of the emission from a localized, rotating and oscillating superluminal source}
\label{IIB}
Consider a localized source (a source whose dimensions are appreciably smaller than
the radiation wavelength, i.e., essentially a point source)
which moves on a circle of radius $r$ with the constant angular velocity
$\omega{\hat{\bf e}}_z$, i.e.\ whose path ${\bf x}(t)$ is given, in terms of the cylindrical polar coordinates $(r,\varphi,z)$, by
$$r={\rm const.},\quad z={\rm const.}, \quad \varphi={\hat\varphi}+\omega t,\eqno(1)$$
where ${\hat{\bf e}}_z$ is the basis vector associated with $z$, and ${\hat\varphi}$ the initial value of $\varphi$.

The wave fronts that are emitted by this point source in an empty and unbounded space are described by
$$\vert{\bf x}_P-{\bf x}(t)\vert=c(t_P-t),\eqno(2)$$
where the coordinates $({\bf x}_P,t_P)=(r_P,\varphi_P,z_P,t_P)$ mark the space-time of observation points.
The distance $\vert{\bf x}_P-{\bf x}\vert\equiv R$ between the observation point ${\bf x}_P$ and the point source ${\bf x}$ is given by
$$R=[(z_P-z)^2+{r_P}^2+r^2-2r_Pr\cos(\varphi_P-\varphi)]^{1\over2},\eqno(3)$$
so that the insertion of Eq.\ (1) in Eq.\ (2) yields
$$t_P=t+[(z_P-z)^2+{r_P}^2+r^2-2r_Pr\cos(\varphi_P-{\hat\varphi}-\omega t)]^{1\over2}/c.\eqno(4)$$
For a given source point $(r, {\hat\varphi}, z)$ and various positions $(r_P, \varphi_P, z_P)$ of
the observation point, this dependence of the reception time $t_P$ on the emission time $t$ can have one of the generic forms shown in Fig.\ 3.

When $r\omega>c$, there is a one-dimensional set of observation points for which $dR/dt=-c$ and
$d^2R/dt^2=0$ at the emission times $t_c$ of the waves, i.e.\ whose members are approached
by the superluminally moving source point with the speed of light and zero acceleration at the retarded time.
These observation points are located on the cusp of the envelope of the emitted wave fronts (Fig.\ 1),
where curve (b) of Fig.\ 3 passes through an inflection point.  In their vicinity, Eq.\ (4) reduces to
$$t_P=t_{Pc}+\textstyle{1\over6}\omega^2(t-t_c)^3+\cdots,\eqno(5)$$
where $t_{Pc}$ is the value of $t_P$ at which the waves emitted at $t=t_c$ arrive, and constructively interfere,
at the cusp curve of the envelope (see Appendix C of~\cite{AAS1}).
Note that the coefficient of the third-order term in the above Taylor expansion happens to be
independent of the coordinates $r$ and $z$ of the source, a feature that enhances the
cooperative (coherent) nature of the process to be described below (Section~\ref{IV}).

\begin{figure}[tbp]
   \centering
\includegraphics[height=7cm]{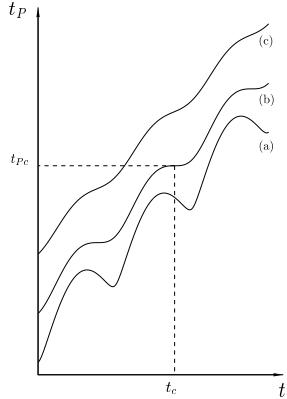}
\caption{The relationship between the observation time $t_P$ and the emission time
$t$ for an observation point that lies (a) inside or on, (b) on the cusp of, and (c)
outside the envelope of the wave fronts or the bifurcation surface shown in Figs.\ 1 and 2.
This relationship is given by $g(r,\varphi,z;r_P,\varphi_P,z_P)={\hat\varphi}-{\hat\varphi}_P$, an equation which applies to the envelope when the position $(r,{\hat\varphi},z)$ of the source point is fixed and to the bifurcation surface when the location $(r_P,{\hat\varphi}_P,z_P)$ of the observer is fixed. (Note that, by virtue of the linear relation $\varphi={\hat\varphi}+\omega t$,
the motion of the source may be parametrized either by $t$ or by ${\varphi}$.)
The maxima and minima of curve (a), at which $dR/dt=-c$, occur on the sheets $\phi_+$ and $\phi_-$
of the envelope (or the bifurcation surface), respectively (see Figs.\ 1 and 2).
The inflection points of curve (b), at which $d^2R/dt^2=0$, occur on the cusp
curve of the envelope (or the bifurcation surface).}
\end{figure}

Now suppose that, in addition to moving faster than its own waves, the point source in question has a strength which
fluctuates with a frequency higher than that of its rotation, like $\cos(\Omega t)$ with $\Omega>\omega$ (Table~1).
The amplitude of the field it would generate will then be proportional to the retarded value of this fluctuating factor,
i.e.\ to a function of $t_P$ that, according to Eq.\ (5), has the form
$\cos\{\Omega[t_c+6^{1\over3}\omega^{-{2\over3}}(t_P-t_{Pc})^{1\over3}]\}$
at points close to the caustics of the wave fronts.
The period of oscillations of the observed field is given by the time interval in which the argument of this
function changes from $\Omega t_c$ to $\Omega t_c+2\pi$.
It will have a value, therefore, that is by the factor ${2\over3}(\pi\omega/\Omega)^2$
shorter than the period $2\pi/\Omega$ of the fluctuations of the source strength.
Stated differently, the time interval $t-t_c=2\pi/\Omega$ in which a set of wave fronts is emitted is
by the factor ${3\over2}(\pi\omega/\Omega)^{-2}$ longer than the time interval $t_P-t_{Pc}$
during which the same set of wave fronts is received.
\subsection{Extension to superluminal volume sources}
\label{IIC}
The field of a uniformly rotating point source constitutes only the
Green's function for the present emission process~\cite{AAS5}.
For a corresponding extended source to emit waves of a certain frequency, it is necessary
in addition that the spectrum of temporal fluctuations of its density should contain that frequency.
In this respect, the frequency enhancing effect associated with a rotating superluminal source
differs radically from that which is familiar from synchrotron radiation.
The space-time distribution of density for the charged particle which acts as the source of synchrotron radiation
entails the Dirac delta function and so has a spectral decomposition that is independent of frequency.
That the maximal intensity in the spectrum of synchrotron radiation corresponds to a frequency which is
much higher than the rotation frequency of its source merely reflects a spectral property of
the Green's function for that emission process (Section~\ref{VII}).
Superluminal sources, on the other hand, are necessarily extended~\cite{AAS1,AAS2,AAS22,AAS23};
the spectra of their densities do not, in general, contain all frequencies.

We now come to a vital distinction between stationary and rotating volume sources which
leads to a radical difference in the spectral content of the associated emission.
This is connected with the differing constraints on the ranges of values of $\varphi$ and ${\hat\varphi}$, constraints which in turn dictate the form assumed by the Fourier decomposition of the source density with respect to time.

The space-time distribution of the rotating {\it point} source described in Eq.\ (1), whose path may
be written as $r=r_0$, $\varphi={\hat\varphi}+\omega t$, $z=0$, has the density
$$\rho(r,\varphi,z,t)=q\,\delta(r-r_0)\delta(\varphi-\omega t-{\hat\varphi})\delta(z)/r,\eqno(6)$$ 
where $\delta$ is the Dirac delta function, $q$ is the volume integral of $\rho$, and $r_0$ a constant.
It is known from the analysis of synchrotron radiation that the azimuthal angle $\varphi$ which appears
in Eqs.\ (1) and (6) is not limited to an interval of length $2\pi$, as in the description of a stationary
source distribution, but (by virtue of having the value ${\hat\varphi}+\omega t$) can range
over the same interval as the time $t$, i.e.\ over ($-\infty,\infty$).
Since $\varphi=2\pi$ represents the same point in space as $\varphi=0$ at any given $(r,z)$,
the source density (6) is periodic, both in $\varphi$ and in $t$.
However, the azimuthal coordinate $\varphi$ does not discontinuously change back to zero each time a rotation is completed.
If the time interval over which the source density is described by Eq.\ (6) exceeds a rotation period, then the angle
$\varphi={\hat\varphi}+\omega t$ that is traversed by the source during this time interval would also exceed $2\pi$.

Now consider a localized {\it volume} source that rotates about the $z$-axis with a
constant angular frequency $\omega$ (Table~1).  The density distribution $\rho(r, \varphi,z,t)$ for a
source of this kind depends on the azimuthal coordinate $\varphi$ in only
the combination ${\hat\varphi}=\varphi-\omega t$, i.e.\ is a function of $(r,{\hat\varphi},z, t)$.
If we label each volume element of this source by the value ${\hat\varphi}$ of its azimuthal coordinate $\varphi$ at $t=0$,
the equations describing the trajectories of these elements would each have the same form as Eq.\ (1).
It can be seen from the collection of space-time trajectories of the constituent volume elements
of this source, therefore, that the ranges of values of both $\varphi$ and $t$ are infinite,
as in the case of a rotating point source, but values of the Lagrangian coordinate
${\hat\varphi}$ are limited to an interval of length $2\pi$, e.g.\ to $-\pi<{\hat\varphi}\le\pi$ (Fig.\ 4).
The coordinate ${\hat\varphi}$ cannot range over a wider interval because no volume element
of an extended source may be labelled by more than one value of a Lagrangian coordinate.
Phrased differently, the aggregate of volume elements that constitute a rotating source
in its entirety can at most occupy an azimuthal interval of length $2\pi$ at any given time (e.g.\ $t=0$).

\begin{figure}[tbp]
   \centering
\includegraphics[height=7cm]{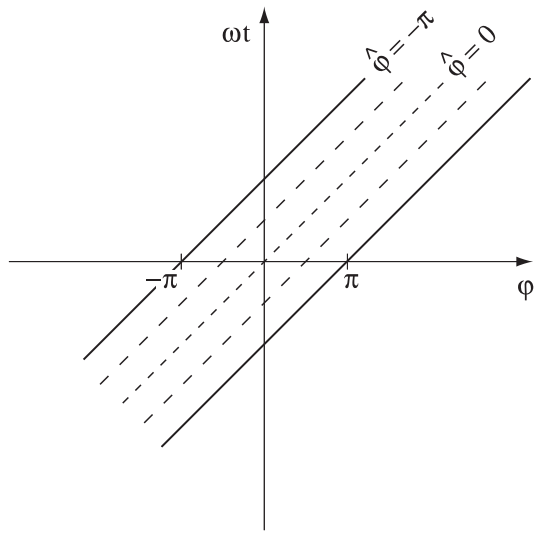}
\caption{Projection of the trajectories (world lines) of the volume elements
of a uniformly rotating extended source onto the $(\varphi, t)$ space.
The Lagrangian coordinate ${\hat\varphi}\equiv\varphi-\omega t$ designating
the initial ($t=0$) position of each source element lies in $(-\pi,\pi)$ while
both $\varphi$ and $t$ range over $(-\infty,\infty)$.  The space-time trajectory
(world tube) of the extended source itself in $(\varphi, t)$ space consists
of the array of trajectories of its constituent volume elements (the broken lines)
encompassed by the lines ${\hat\varphi}=-\pi$ and ${\hat\varphi}=\pi$.}
\end{figure}

\begin{figure}[tbp]
   \centering
\includegraphics[height=5cm]{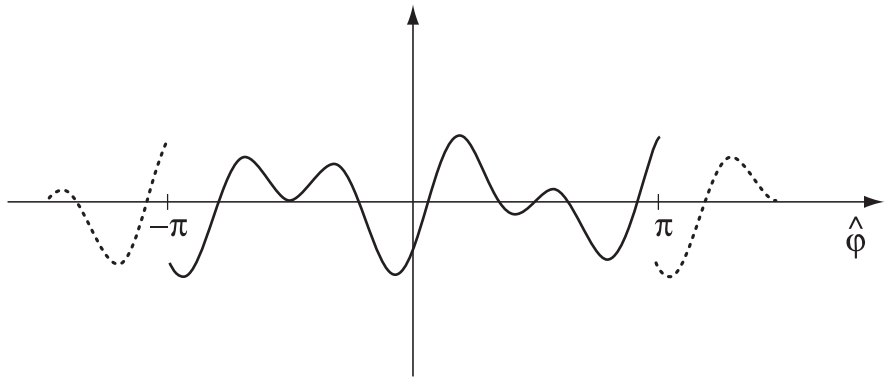}
\caption{The ${\hat\varphi}$ dependence, $\cos(m{\hat\varphi})\cos[\Omega({\hat\varphi}-\varphi)/\omega]$,
of the source density decribed by Eq.\ (7) at a fixed
$(r,\varphi,z)$ for $m=1$, $\Omega/\omega=3.5$ and $\varphi=\pi/5$.
This ${\hat\varphi}$ dependence is, by virtue of the relationship $t=({\hat\varphi}-\varphi)/\omega$,
equivalent to a time dependence:  the space-time of source elements may be marked
either by the coordinates $(r,\varphi,z,t)$, as in Eq.\ (7), or by the coordinates
$(r,\varphi,z,{\hat\varphi})$ as in Eq.\ (19).  The plotted function is
physically meaningful only within the interval $-\pi<{\hat\varphi}<\pi$.
However, once this function is expanded into a Fourier series over the interval $(-\pi,\pi)$,
a new periodic function results, represented by the series, which is coincident with the
original function within $(-\pi,\pi)$ and periodically reproduces the original function outside this interval.
The periodic function represented by the series outside $(-\pi,\pi)$ is designated by broken lines.
That the Fourier expansion of source density (7) should contain an infinite number of terms may be
understood in two (mathematically equivalent) ways:  either as due to the discontinuities
of the periodic function shown here or as due to the non-zero difference between
the values of this density at ${\hat\varphi}=\pm\pi$.}
\end{figure}

\subsection{Algebraic representation of polarization currents with superluminally rotating distribution patterns}
\label{IID}
For an extended source of radiation whose distribution pattern rotates uniformly,
the cylindrical components of the electric current density, ${\bf j}$, similarly depend
on $\varphi$ only via the Lagrangian coordinate ${\hat\varphi}\equiv\varphi-\omega t$:
they have the space-time dependence $j_{r,\varphi,z}(r, \varphi, z, t)=j_{r,\varphi,z}(r, {\hat\varphi}, z)$,
where $j_{r,\varphi,z}$ stand for the components of ${\bf j}$ along the
cylindrical base vectors $({\hat{\bf e}}_r, {\hat{\bf e}}_\varphi, {\hat{\bf e}}_z)$.
In this paper we consider sources that oscillate in addition to moving,
i.e.\ sources for which $j_{r,\varphi,z}$ are given by $j_{r,\varphi,z}(r,{\hat\varphi},z)f(t)$ with an additional
dependence $f(t)$ on time.
To be specific, we base the analysis on a representative polarization current ${\bf j}=\partial{\bf P}/\partial t$ for which
$$P_{r,\varphi,z}(r, \varphi, z, t)=s_{r,\varphi,z}(r, z)\cos(m{\hat\varphi})\cos(\Omega t),\qquad -\pi<{\hat\varphi}\leq\pi,\eqno(7)$$
where $P_{r,\varphi,z}$ are the cylindrical components of the polarization
(the electric dipole moment per unit volume), ${\bf s}(r,z)$ is an arbitrary vector
that vanishes outside a finite region of the $(r, z)$ space and $m$ is a positive integer (Table~1).

For a fixed value of $t$, the azimuthal dependence of the above density along each circle of radius $r$
within the source is the same as that of a sinusoidal wave train, with the wavelength $2\pi r/m$,
whose $m$ cycles fit around the circumference of the circle smoothly.
As time elapses, this wave train both propagates around each circle with the velocity
$r\omega$ and oscillates in its amplitude with the frequency $\Omega$ (Table~1).
The vector ${\bf s}$ is here left arbitrary in order that we may
later investigate the polarization
of the resulting radiation for all possible directions of the emitting current
(this will be summarised in Table~2).
Note that one can construct any distribution with a uniformly rotating pattern,
$P_{r,\varphi,z}(r, {\hat\varphi}, z)$, by the superposition over $m$
of terms of the form $s_{r,\varphi,z}(r, z, m)\cos(m{\hat\varphi})$.

An experimentally viable device capable of generating a polarization
with the distribution (7) is described in Appendix A~\cite{AAS12,johan}.
Even though the practical implementation of the polarization (7) by means of this specific device
only entails the two frequencies $m\omega$ and $\Omega$, the spectral decomposition of this
polarization consists of an infinite set of frequencies if $\Omega/\omega$ is different from an integer.
One can directly demonstrate this by Fourier analyzing the right-hand side of Eq.\ (7).

Because the domain of definition of the source density (7) only extends over
the interval $-\pi<{\hat\varphi}\leq\pi$, or equivalently $(\varphi-\pi)/\omega\le t<(\varphi+\pi)/\omega$,
Fourier decomposition of the time dependence of this function at a given $(r,\varphi,z)$
should be performed by means of a series rather than an integral.
Representation of Eq.\ (7) by a Fourier integral would entail assumptions about
the dependence of $j_{r,\varphi,z}$ on ${\hat\varphi}$ in intervals
($-\infty<{\hat\varphi}<-\pi$ and $\pi<{\hat\varphi}<\infty$) which lie outside the domain
of definition of this density.  The following Fourier series faithfully represents the right-hand side of
Eq.\ (7) within its domain of definition and replaces it by a periodic function outside the physically
relevant domain $-\pi<{\hat\varphi}\leq\pi$:
$$P_{r,\varphi,z}(r,\varphi,z,t)=s_{r,\varphi,z}(r,z)\sum_{n=-\infty}^\infty C_n\exp ({\rm i}n\omega t),\eqno(8)$$
where 
%$$\eqalignno{
\begin{eqnarray*}
C_n=&(2\pi/\omega)^{-1}\int_{(\varphi-\pi)/\omega}^{(\varphi+\pi)/\omega}dt\,\cos[m(\varphi-\omega t)]\cos(\Omega t)\exp(-{\rm i}n\omega t)\cr
&=(-1)^{m+n+1}(2\pi)^{-1}(n-\Omega/\omega)[(n-\Omega/\omega)^2-m^2]^{-1}\qquad\quad\cr
&\times\sin(\pi\Omega/\omega)\exp[-{\rm i}(n-\Omega/\omega)\varphi]+\{\Omega\to -\Omega\}.\qquad&(9)%\cr
\end{eqnarray*}
%}$$
The symbol $\{\Omega\to -\Omega\}$ designates a term like the one preceding
it in which $\Omega$ is everywhere replaced by $-\Omega$.

Note that the Fourier components $C_n$ of the function describing the time dependence of
$P_{r,\varphi,z}$ would vanish for all $n\neq(\Omega/\omega)\pm m$ only if $\Omega/\omega$
is an integer (Table 1).
When $\Omega$ and $\omega$ are not commensurable, the factor $\sin(\pi\Omega/\omega)$
in Eq.\ (9) is different from zero and the spectrum of $P_{r,\varphi,z}$ contains all frequencies.
It is not difficult to see the reason for this:  the Fourier series representation of $P_{r,\varphi,z}$
equals a periodic function in $-\infty<{\hat\varphi}<\infty$ whose values at the beginning and
at the end of a period are different when $\Omega/\omega$ is different from an integer (see Fig.\ 5).
The higher frequencies stem from these step-like discontinuities of the global function
represented by the series, discontinuities which lie outside the physically relevant domain
$-\pi<{\hat\varphi}\le\pi$ but which nevertheless mathematically influence
the Fourier expansion of the limited part of this function that describes the source density.
\subsection{The role of centripetal acceleration in providing broadband emission}
\label{IIE}
Physically, the agent responsible for this implicit discontinuity and so the broadening
of the spectrum of the source density is centripetal acceleration.
That the spectral content of a {\it rectilinearly} moving source is broadened
by acceleration can be directly seen from the transformation $x^\prime=x+ut+{1\over2}at^2$
between an accelerated frame $S$ (with the initial velocity $u$ and acceleration $a$) and an inertial frame $S^\prime$.
A source that is spatially monochromatic in its own rest frame, e.g.\ $\exp({\rm i}kx)$
with the wavenumber $k$, is transformed into one, $\exp[{\rm i}k(x^\prime-ut-{1\over2}at^2)]$,
whose spectrum does not even decay at high frequencies.
The corresponding effect of centripetal acceleration on the spectrum of a
source is more subtle and is manifested in a kinematic constraint set by the geometry of rotation.

Because there is only one parameter ($\omega$) for describing both the speed ($r\omega$)
and the acceleration ($r\omega^2$) of a uniformly rotating source element, the centripetal
acceleration of a rotating extended source shows up in the constraint $-\pi<{\hat\varphi}\leq\pi$
on the range of values of the Lagrangian coordinate ${\hat\varphi}$:
an aspect of the geometry of rotation that acts as an additional parameter.
Had the range of ${\hat\varphi}$ been infinite, the time dependence of the distribution in
Eq.\ (7) would have been indistinguishable from that of a rectilinearly moving source
with a constant velocity and its Fourier transform would have contained only the
frequencies $\vert\Omega\pm m\omega\vert$, irrespective of whether $\Omega$ and $\omega$
are commensurable or not.
It is the fact that ${\hat\varphi}$ is an angle marking the elements
of a rotating source that gives rise to the constraint $-\pi<{\hat\varphi}\leq\pi$,
to effective space-time discontinuities in the source density, and to the higher frequencies.

Given that both a representative source density and the Green's function for the present emission
process have spectra which contain infinite sets of frequencies, it is not unexpected that the
analysis in Sections~\ref{V} and \ref{VI} should predict a resulting radiation that is correspondingly broadband.
This prediction is not incompatible with the fact that oscillations at no more than
two frequencies ($m\omega$ and $\Omega$) are required for creating the representative
source distribution [Eq.\ (7)] in the laboratory (Appendix A; Table~1).
As in the case of any other linear system, the present emission process generates an
output only at those frequencies which are carried both by its input (the source) and its response (Green's) function.
What entails only two frequencies is the practical implementation of the source
we are considering and not its spectral content.
The density distribution of the present source includes implicit space-time discontinuities
(Fig.\ 5) whose Fourier decompositions contain all frequencies [Eq.\ (9)].
These step-like discontinuities do not require (for their practical implementation)
the creation of any rapid changes either in space or in time; they have to do
with the geometry of rotation and automatically stem from centripetal acceleration.

The radiation field that arises from the superluminal portion ($r>c/\omega$) of the volume source
described in Eq.\ (7) consists (as shown in~\cite{AAS1}) of two components:
a spherically decaying component whose intensity diminishes like ${R_P}^{-2}$
with the distance $R_P$ from the source, and a non-spherically spreading component
whose intensity diminishes more slowly with distance.
The analysis in this paper is concerned only with the spectral properties of the spherically decaying
component of the radiation.  The component of the radiation whose intensity
decays like ${R_P}^{-1}$ instead of ${R_P}^{-2}$ is emitted
only at the two frequencies $\vert\Omega\pm m\omega\vert$~\cite{newpaper}.
\section{Detailed formulation of the problem}
\label{III}
\subsection{Electromagnetic fields in the far-field limit}
\label{IIIA}
In the absence of boundaries, the retarded potential $A^\mu$ arising from any localized
distribution of charges and currents with a density $j^\mu$ is given by
$$A^\mu({\bf x}_P,t_P)=c^{-1}\int d^3x\,dt\,j^\mu({\bf x},t)\delta(t_P-t-R/c)/R,\quad \mu=0,\cdots,3\eqno(10)$$
where $R$ stands for the magnitude of ${\bf R}\equiv{\bf x}_P-{\bf x}$, and $\mu=1, 2, 3$
designate the spatial components, ${\bf A}$ and ${\bf j}$, of $A^\mu$ and $j^\mu$ in a Cartesian coordinate system.
The expressions that follow for the electromagnetic fields
$${\bf E}=-{\bf\nabla}_PA^0-\partial{\bf A}/\partial(ct_P)\qquad{\rm and}\qquad {\bf B}={\bf\nabla}_P{\bf\times A}\eqno(11)$$
when we simply differentiate Eq.\ (10) under the integral sign, and evaluate the resulting integrals by parts, are
$${\bf E}=-{1\over c}\int d^3x\,dt\,{\delta(t_P-t-R/c)\over R}\Big[{1\over c}{\partial{\bf j}\over\partial t}-\Big({1\over c}{\partial j^0\over\partial t}+{j^0\over R}\Big){{\bf R}\over R}\Big],\eqno(12)$$
$${\bf B}=-{1\over c}\int d^3x\,dt\,{\delta(t_P-t-R/c)\over R}{{\bf R}\over R}{\bf\times}\Big({1\over c}{\partial{\bf j}\over\partial t}+{{\bf j}\over R}\Big),\eqno(13)$$
since $j^\mu$ vanishes outside a finite volume.

Terms of the order of $R^{-2}$ in the above integrands, which do not contribute toward the
flux of energy at infinity, may be discarded if we are concerned only with the radiation field.
Since the problem we will be considering entails the formation of caustics, however,
we need to treat the phases of the above integrands, i.e.\ the arguments of the delta
functions in Eqs.\ (12) and (13), more accurately.
If we replace $\partial j^0/\partial(ct)$ in Eq.\ (12) with $-{\bf\nabla\cdot j}$ from continuity,
integrate this term by parts and retain only those terms in the integrands of Eqs.\ (12) and (13) that are of the order of $R^{-1}$, we obtain
$${\bf E}\simeq{1\over c^2}\int d^3x\,dt\,{\delta(t_P-t-R/c)\over R}{\hat{\bf n}}{\bf\times}\Big({\hat{\bf n}}{\bf\times}{\partial{\bf j}\over\partial t}\Big)\eqno(14)$$
and ${\bf B}\simeq{\hat{\bf n}}{\bf\times}{\bf E}$.
Here, we have set the origin of the coordinate system within the source distribution so that
$\vert{\bf x}\vert\ll\vert{\bf x}_P\vert$ for an observation point in the far field and ${\bf R}/R$
can be approximated by the constant vector ${\hat{\bf n}}\equiv{\bf x}_P/\vert{\bf x}_P\vert$; the symbol $\simeq$ indicates that the expression is valid in the far-field limit.
These differ from the standard expressions for radiation fields~\cite{AAS11} only in that the argument of the delta function in their integrands is left exact.

For the purposes of evaluating the integrals in Eq.\ (14) for the current density that is given by Eq.\ (7),
the space-time of source points may be marked either with $({\bf x},t)=(r,\varphi,z,t)$
or with the coordinates $(r,{\hat\varphi},z,t)$ that naturally appear in the description of that rotating source.
In fact, once ${\hat\varphi}$ is adopted as the coordinate that ranges over $(-\pi,\pi)$, the retarded
position $\varphi$ of the rotating source point $(r, {\hat\varphi},z)$ as well as the retarded time
$t$ could be used as the coordinate whose range is unlimited (see Fig.\ 4).

The electric current density ${\bf j}=\partial{\bf P}/\partial t$ that arises from the
polarization distribution (7) is given, in terms of $\varphi$ and ${\hat\varphi}$, by
%$$\eqalignno{
\begin{eqnarray*}
{\bf j}=&\textstyle{1\over4}{\rm i}\omega\sum_{\mu=\mu_\pm}\mu\exp[-{\rm i}(\mu{\hat\varphi}-\Omega\varphi/\omega)](s_r{\hat{\bf e}}_r+s_\varphi{\hat{\bf e}}_\varphi+s_z{\hat{\bf e}}_z)\cr
&+\{m\to-m,\Omega\to-\Omega\},\quad-\pi<{\hat\varphi}\leq\pi,&(15)%\cr
\end{eqnarray*}
%}$$
where $\mu_\pm\equiv(\Omega/\omega)\pm m$ and the symbol $\{m\to-m,\Omega\to-\Omega\}$
designates a term like the one preceding it in which $m$ and $\Omega$ are everywhere
replaced by $-m$ and $-\Omega$, respectively.
To put this source density into a form suitable for inserting in Eq.\ (14), we need to express the
$\varphi$-dependent base vectors $({\hat{\bf e}}_r, {\hat{\bf e}}_\varphi, {\hat{\bf e}}_z)$
associated with the source point $(r, \varphi, z)$ in terms of the constant base vectors
$({\hat{\bf e}}_{r_P},{\hat{\bf e}}_{\varphi_P}, {\hat{\bf e}}_{z_P})$ at the observation point $(r_P, \varphi_P,z_P)$:
$$\left[\matrix{{\hat{\bf e}}_r\cr {\hat{\bf e}}_\varphi\cr {\hat{\bf e}}_z\cr}\right]=\left[\matrix{\cos(\varphi-\varphi_P)&\sin(\varphi-\varphi_P)&0\cr
-\sin(\varphi-\varphi_P)&\cos(\varphi-\varphi_P)&0\cr
0&0&1\cr}\right]\left[\matrix{{\hat{\bf e}}_{r_P}\cr {\hat{\bf e}}_{\varphi_P}\cr {\hat{\bf e}}_{z_P}\cr}\right].\eqno(16)$$
Equations (15) and (16) together with the far-field value of ${\hat{\bf n}}$,
$$\lim_{R\to\infty}{\hat{\bf n}}=\sin\theta_P{\hat{\bf e}}_{r_P}+\cos\theta_P{\hat{\bf e}}_{z_P},\quad\theta_P\equiv\arctan(r_P/z_P),\eqno(17)$$
yield the following expression for the source term in Eq.\ (14):
%$$\eqalignno{
\begin{eqnarray*}
{\hat{\bf n}}{\bf\times}({\hat{\bf n}}{\bf\times}\partial{\bf j}/\partial t)=&\textstyle{1\over4}\omega^2\sum_{\mu=\mu_\pm}\mu^2\exp[-{\rm i}(\mu{\hat\varphi}-\Omega\varphi/\omega)]\big\{[s_\varphi\cos\theta_P\sin(\varphi-\varphi_P)\cr
&-s_r\cos\theta_P\cos(\varphi-\varphi_P)+s_z\sin\theta_P]{\hat{\bf e}}_\perp+[s_r\sin(\varphi-\varphi_P)\cr
&+s_\varphi\cos(\varphi-\varphi_P)]{\hat{\bf e}}_\parallel\big\}+\{m\to-m,\Omega\to -\Omega\},\qquad&(18)
\end{eqnarray*}
%\cr}$$
where ${\hat{\bf e}}_\parallel\equiv {\hat{\bf e}}_{\varphi_P}$ (which is parallel to the plane of rotation)
and ${\hat{\bf e}}_\perp\equiv{\hat{\bf n}}{\bf\times}{\hat{\bf e}}_\parallel$ comprise a pair of unit vectors normal to the radiation direction ${\hat{\bf n}}$.
\subsection{Green's functions}
\label{IIIB}
Inserting Eq.\ (18) in Eq.\ (14) and changing the variables of integration from $({\bf x}, t)=(r, \varphi, z, t)$ to $(r, \varphi, z, {\hat\varphi})$, we obtain
%$$\eqalignno{
\begin{eqnarray*}
{\bf E}\simeq&\textstyle{1\over4}(\omega/c)^2\sum_{\mu=\mu_\pm}\int_0^\infty r\, dr\int_{-\infty}^{+\infty}dz\int_{-\pi}^{+\pi}d{\hat\varphi}\,\mu^2\exp(-{\rm i}\mu{\hat\varphi})\big\{(s_r G_2+s_\varphi G_1){\hat{\bf e}}_\parallel\cr
&+[\cos\theta_P(s_\varphi G_2-s_r G_1)+\sin\theta_P s_z G_3]{\hat{\bf e}}_\perp\big\}+\{m\to-m,\Omega\to -\Omega\},\quad&(19)
%\cr}$$
\end{eqnarray*}
where $\simeq$ again indicates that the expression is accurate in the far-field limit, and
$G_i$ ($i=1,2,3$) are the functions resulting from the remaining integration with respect to $\varphi$:
$$\left[\matrix{G_1\cr G_2\cr G_3\cr}\right]=\int_{\Delta\varphi} d\varphi\,{\delta(g-\phi)\over R}\exp({\rm i}\Omega\varphi/\omega)\left[\matrix{\cos(\varphi-\varphi_P)\cr \sin(\varphi-\varphi_P)\cr 1\cr}\right].\eqno(20)$$
Here $R$ is as in Eq.\ (3), $\phi$ stands for ${\hat\varphi}-{\hat\varphi}_P$ with ${\hat\varphi}\equiv \varphi-\omega t$ and ${\hat\varphi}_P\equiv \varphi_P-\omega t_P$, the function $g$ is defined by
$$g\equiv\varphi-\varphi_P+{\hat R},\eqno(21)$$
with ${\hat R}\equiv R\omega/c$, and $\Delta\varphi$ is the interval of azimuthal angle traversed by the source.
The $r$ and $z$ integrations in Eq.\ (19), though extending over the entire $(r,z)$ space,
of course receive contributions only from those regions of this space in which the source densities $s_{r,\varphi,z}$ are non-zero.

Note that it would make no difference to the outcome of the calculation whether one uses the expression in
Eq.\ (19) and integrates over the coordinates $({\hat\varphi},\varphi)$ as we have done, or
one uses Eq.\ (14) with $({\bf x}, t)=(r,\varphi,z,t)$ and integrates over $(t,\varphi)$ as is conventionally done.
If one follows the conventional procedure and first integrates with respect to $t$, then
the constraint $-\pi<{\hat\varphi}\le\pi$ would show up as the restriction
$(\varphi-\pi)/\omega\le t<(\varphi+\pi)/\omega$ on the range of $t$ integration (see Fig.\ 4).

The functions $G_i(r,{\hat\varphi}, z; r_P,{\hat\varphi}_P, z_P, \varphi_P)$ here act as Green's
functions:  they describe the fields of uniformly rotating point sources with fixed (Lagrangian)
coordinates $(r,{\hat\varphi}, z)$ whose strengths sinusoidally vary with time.
The field (19) is given by the superposition of the fields of the assembly of such uniformly rotating volume elements
from which the extended source (15) is built up.
In the special case in which $\Omega=0$, i.e.\ the strength of the source is constant,
$G_3$ reduces to the Green's function called $G_0$ in~\cite{AAS1}.
The singularity structures of $G_i$ are determined by the stationary points of the phase function $g$
and so are identical to the singularity structure already outlined in connection with $G_0$ (see~\cite{AAS1}).
\subsection{Spectral decomposition of the radiated field}
\label{IIIC}
Spectral decomposition of the radiated field ${\bf E}$ may be achieved, as in any other
time-dependent problem, simply by replacing the delta function in Eq.\ (20) with its Fourier representation.
Because the integration with respect to ${\hat\varphi}$ only extends over the interval $(-\pi,\pi)$ (Section~\ref{IIC}),
Fourier decomposition of the ${\hat\varphi}$ dependence of this delta function
should be performed by means of a series.

The integrand in Eq.\ (19) needs to be faithfully represented only within the range of integration.
Representation of this integrand by a Fourier integral would entail assumptions about the dependence of
$j_{r,\varphi,z}$ on ${\hat\varphi}$ in intervals which lie outside the domain of
definition of the source density:  in $-\infty<{\hat\varphi}<-\pi$ and $\pi<{\hat\varphi}<\infty$ (see Section~\ref{II}).

Once the delta function $\delta(g-\phi)$ that appears in the integral representation of
$G_i$ in Eq.\ (20) is expanded into a Fourier series over the interval $-\pi<{\hat\varphi}<\pi$,
$$\delta(g-\phi)=(2\pi)^{-1}\sum_{n=-\infty}^\infty\exp[-{\rm i}n(g-\phi)],\eqno(22)$$
Eq.\ (19) becomes
$${\bf E}=\Re\Big\{{\tilde{\bf E}}_0+2\sum_{n=1}^{\infty}{\tilde{\bf E}}_n\exp(-{\rm i}n{\hat\varphi}_P)\Big\}, \eqno(23)$$
in which
%$$\eqalignno{
\begin{eqnarray*}
{\tilde{\bf E}}_n\simeq&\textstyle{1\over4}(\omega/c)^2\sum_{\mu=\mu_\pm}\int_0^\infty r\, dr\int_{-\infty}^{+\infty}dz\int_{-\pi}^{+\pi}d{\hat\varphi}\,\mu^2\exp(-{\rm i}\mu{\hat\varphi})\big\{(s_r\tilde{G_2}+s_\varphi\tilde{G_1}){\hat{\bf e}}_\parallel\cr
&+[\cos\theta_P(s_\varphi\tilde{G_2}-s_r\tilde{G_1})+\sin\theta_P s_z\tilde{G_3}]{\hat{\bf e}}_\perp\big\}+\{m\to-m,\Omega\to -\Omega\},\quad&(24)
\end{eqnarray*}
%\cr}$$
and $\Re\{Z\}$ stands for the real part of $Z$.  The functions $\tilde{G}_i$ in this expression are given by
$${\tilde{G}}_i=(2\pi)^{-1}\exp({\rm i}n{\hat\varphi})\int_{\Delta\varphi}d\varphi f_i\exp[-{\rm i}(ng-\Omega\varphi/\omega)],\eqno(25)$$
with
$$\left[\matrix{f_1\cr f_2\cr f_3\cr}\right]\equiv R^{-1}\left[\matrix{\cos(\varphi-\varphi_P)\cr \sin(\varphi-\varphi_P)\cr 1\cr}\right],\eqno(26)$$
and constitute the Fourier components of the Green's functions $G_i$.

Because the values of ${\hat\varphi}$ are limited to an interval of length $2\pi$, the radiation
is emitted in harmonics $n\omega$ of the rotation frequency (Table~1).
The periodic nature of the motion of the source imposes this constraint despite the fact that the source distribution
[Eq.\ (7)] lacks periodicity.
[Recall that the ratio $\Omega/\omega$ that appears in the expression for the source
density, Eq.\ (7), is different from an integer.]  In the regime $\Omega/\omega\gg1$,
however, the peak of the spectrum happens to occur at such a high value of the harmonic number
$n$ that this spectrum is essentially continuous.
\section{Loci of coherently-contributing source elements}
\label{IV}
\subsection{The importance of focal regions in the space of observation points}
\label{IVA}
The filamentary cusps of the envelopes of the wave fronts that emanate from various volume elements of an extended superluminal source (Fig.\ 1) collectively occupy a tubular volume of the ${\bf x}_P$ space, a volume which we shall refer to as the focal region of the space of observation points.  At any given observation point $P$ within this focal region, there are certain volume elements of the
source whose contributions towards the value of the field at the observation time $t_P$
superpose coherently, i.e.\ arrive at ${\bf x}_P$ with the same phase.
These consist of those elements of the superluminally moving source which approach
the observer along the radiation direction with the speed of light and zero
acceleration at the retarded time (Section~\ref{IIB}).
Or stated mathematically, for large values of the harmonic number $n$,
the main contributions towards the value of the multiple integral (24) representing
the radiation field $\tilde{\bf E}_n$ come from the stationary points of the optical distance
$t_P-t-R/c$, given by the function $g(r,\varphi,z)$ of Eq.\ (21), that appears in the
phase of the rapidly oscillating exponential $\exp(-{\rm i}ng)$ in the integrand of this integral~\cite{AAS7,AAS72,AAS8}.
As a first step towards the asymptotic evaluation of the multiple integral in Eq.\ (24),
therefore, we need to identify the loci of points at which the derivatives
$\partial g/\partial r$, $\partial g/\partial\varphi$ and $\partial g/\partial z$
vanish and to expand $g$ into a Taylor series about each of its stationary points.

There is in the present case a point at which all three of the above derivatives are zero.
The coordinates of this point, which we shall designate as $O$, are given by
$$O:\quad{\hat r}=1,\quad \varphi=\varphi_P+2\pi-\arccos(1/{\hat r}_P)\equiv\varphi_O,\quad z=z_P,\eqno(27)$$
where ${\hat r}\equiv r\omega/c$ and ${\hat r}_P\equiv r_P\omega/c$ stand
for the values of $r$ and $r_P$ in units of the light-cylinder radius $c/\omega$.
The second derivative $\partial^2 g/\partial\varphi^2$ of $g$ also vanishes at $O$.
Since the derivative of $g$ with respect to $\varphi$ at fixed $(r,{\hat\varphi},z)$ is
proportional to the derivative of $t-t_P+R/c$ with respect to $t$ at fixed $(r,\varphi,z)$,
the conditions $\partial g/\partial\varphi=0$ and $\partial^2 g/\partial\varphi^2=0$ respectively
correspond to the conditions $dR/dt=-c$ and $d^2R/dt^2=0$ which we encountered in Section~\ref{IIB} (see Fig.\ 3).
Not only does the point $O$ belong to the locus of source points that approach the observer with
the speed of light and zero acceleration at the retarded time
(i.e.\ lies on the cusp curve of the observer's bifurcation surface (Figs. 1 and 2,~\cite{AAS1}),
but it is in fact the point at which this locus touches, and is tangential to, the light cylinder
$r=c/\omega$ (see Fig.\ 2).

Whether or not the above stationary point falls within the domain of integration depends
on the position of the observation point.
For a localized source distribution whose dimensions are much smaller than the distance of the
observer from the source, $O$ would fall within the domain of integration
only if the observer lies in the plane of rotation, i.e.\ if the
plane which passes through the observation point and is normal to the rotation axis
intersects the source distribution; otherwise, there would be no source points for which $z$ equals $z_P$.

For an observer who is located outside the plane of rotation, i.e.\ whose coordinate $z_P$ does not match the
coordinate $z$ of any source element, only $\partial g/\partial r$ and $\partial g/\partial\varphi$ can vanish
simultaneously.  This occurs along the curve
%$$\eqalignno{
\begin{eqnarray*}
C:\quad&{\hat r}={\hat r}_C({\hat z})\equiv\{\textstyle{1\over2}({{\hat r}_P}^2+1)-[\textstyle{1\over4}({{\hat r}_P}^2-1)^2-({\hat z}-{\hat z}_P)^2]^{1\over2}\}^{1\over2},\cr
&\varphi=\varphi_C({\hat z})\equiv\varphi_P+2\pi-\arccos({\hat r}_C/{\hat r}_P).\quad&(28)
\end{eqnarray*}
%\cr}$$
In the far-field limit, where the terms $({\hat z}-{\hat z}_P)^2/({{\hat r}_P}^2-1)^2$ and ${\hat r}_C/{\hat r}_P$ in Eq.\ (28)
are much smaller than unity, this curve coincides with the locus
%$$\eqalignno{
\begin{eqnarray*}
C_b:\quad&{\hat r}=[1+({\hat z}-{\hat z}_P)^2/({{\hat r}_P}^2-1)]^{1\over2},\cr
&\varphi=\varphi_P+2\pi-\arccos[1/({\hat r}{\hat r}_P)],\qquad&(29)
\end{eqnarray*}
%\cr}$$
of source points which approach the observer along the radiation direction with the wave speed and zero acceleration
at the retarded time, i.e.\ it coincides with the cusp curve of the bifurcation surface (Fig.\ 2).
\subsection{The inadequacy of conventional far-field approximations}
\label{IVB}
The above calculation makes it clear how essential it is that one should start with the exact form of the optical distance
$\vert{\bf x}_P-{\bf x}(t)\vert-c(t_P-t)$ for identifying the loci of its stationary points.
The far-field approximation $\vert{\bf x}_P-{\bf x}(t)\vert\simeq \vert{\bf x}_P\vert-{\bf x}\cdot{\bf x}_P/\vert{\bf x}_P\vert$
that is normally introduced at the outset of a calculation in radiation theory~\cite{AAS11} here would obliterate,
not only significant geometrical features of the loci of these stationary points,
but also such determining characteristics as the degree of their degeneracy.
The far-field approximation would replace the function $g$ by
$$g\simeq{\hat R}_P-{\hat z}\cos\theta_P+\varphi-\varphi_P-{\hat r}\sin\theta_P\cos(\varphi-\varphi_P),\quad{\hat R}_P\gg1,\eqno(30)$$
where ${\hat R}_P\equiv({{\hat r}_P}^2+{{\hat z}_P}^2)^{1\over2}$.
It is hardly possible to discern the geometrical details of $O$ and $C$ from this expression, let alone their nature.

To preserve the essential features of $g$ about its critical point $O$, we need to express this function in terms of the variables
$$\xi\equiv\varphi-\varphi_O,\quad\eta\equiv{\hat r}-1,\quad\zeta\equiv{\hat z}-{\hat z}_P,\eqno(31)$$
prior to proceeding to the far-field limit.  The resulting exact expression for $g(\xi,\eta,\zeta)$ reduces, when expanded in powers of ${{\hat r}_P}^{-1}$, to
$$g=\phi_O+\xi-(1+\eta)\sin\xi+[4(1+\eta)\sin^2(\xi/2)-(1+\eta)^2\sin^2\xi+\eta^2+\zeta^2]/(2{\hat r}_P)+\cdots,\eqno(32)$$
where $\phi_O\equiv{\hat r}_P+\varphi_O-\varphi_P$.  The term of order ${{\hat r}_P}^{-1}$
in this expansion clearly plays a crucial role in determining the nature of the stationary point $\xi=\eta=\zeta=0$
and so cannot be discarded as in conventional radiation theory.
For the purposes of calculating the asymptotic values of the radiation integrals by the method of stationary phase~\cite{AAS7,AAS72,AAS8},
however, it is mathematically permissible to approximate the coefficient of this term by means of a
Taylor expansion about $O$.  To within the third order in $\xi$ and $\eta$, the result is $(\eta^2+\zeta^2)/2$ so that
$$g=\phi_O+\xi-(1+\eta)\sin\xi+(\eta^2+\zeta^2)/(2{\hat r}_P)+\cdots.\eqno(33)$$
This is a more accurate version of the far-field approximation which, in contrast to that appearing in Eq.\ (30),
exhibits the nondegenrate nature of the stationary point $O$ explicitly:
neither the coefficient of $\eta^2$ nor that of $\zeta^2$ are zero in the Taylor expansion of $g$ about $O$.

The corresponding expansion of $g$ about a point $({\hat r}_C, \varphi_C, {\hat z})$ on curve $C$
(with an arbitrary coordinate ${\hat z}\neq{\hat z}_P$) can likewise be found by first rewriting this function in terms of the variables
$$\rho\equiv{\hat r}-{\hat r}_C,\quad\mu\equiv\varphi-\varphi_C.\eqno(34)$$  
The result is
$$g=\varphi_C-\varphi_P+\mu+[{{\hat R}_C}^2-2(1+\rho/{\hat r}_C){\hat R}_C\sin\mu+4{\hat r}_C({\hat r}_C+\rho)\sin^2(\mu/2)+\rho^2]^{1\over2},\eqno(35)$$
in which we have denoted the value of ${\hat R}$ on $C$ by 
$${\hat R}_C\equiv[({\hat z}_P-{\hat z})^2+{{\hat r}_P}^2-{{\hat r}_C}^2]^{1\over2}\eqno(36)$$
and have made use of the fact that, according to Eqs.\ (28) and (36), ${\hat r}_C({{\hat r}_P}^2-{{\hat r}_C}^2)^{1\over2}$ equals ${\hat R}_C$.\par
If we now expand the right-hand side of Eq.\ (35) in powers of ${{\hat R}_C}^{-1}$
(which tends to ${{\hat R}_P}^{-1}$ in the far zone) and approximate the coefficient of
${{\hat R}_C}^{-1}$ in the resulting expansion by its value in the vicinity of $\rho=\mu=0$, we arrive at
$$g=\phi_C+\mu-(1+\rho/{\hat r}_C)\sin\mu+[\rho^2+({{\hat r}_C}^2-1)\mu^2]/(2{\hat R}_C)+\cdots,\eqno(37)$$
where $\phi_C\equiv{\hat R}_C+\varphi_C-\varphi_P$.
The term $\rho^2$ in the coefficient of ${{\hat R}_C}^{-1}$ in Eq.\ (37) plays an
essential role in specifying the degree to which $g$ is stationary on $C$:  $\rho$
only appears linearly in the earlier terms of the expansion.
The term $\mu^2$ in this coefficient, on the other hand, merely represents a small correction of the order of
${{\hat R}_C}^{-1}$ to the dependence of the value of $g$ on $\mu$ and is of no
consequence as far as the asymptotic values of the radiation integrals are concerned.
We shall therefore neglect the term ${{\hat R}_C}^{-1}\mu^2$ in Eq.\ (37) from now on.

We have already alluded to the distinction between observation points in and outside the plane of the source (Section~\ref{IVA});
we shall now treat these two cases separately in Sections~\ref{V} and \ref{VI}.
\section{The radiation field in the plane of the source's orbit}
\label{V}
\subsection{Treatment of individual volume elements; asymptotic expansion of the Green's function}
\label{VA}
Suppose that the observation point is located within the region $z_<\le z_P\le z_>$ spanned by the orbital planes of
various volume elements of the source and that its coordinate $\varphi_P$ at the observation time $t_P$ is such
that the stationary point $O$ falls within the range $\Delta\varphi$ of integration in Eq.\ (25) ($z_<<0$ and $z_>>0$
stand for the extremities of the $z$ extent of the source distribution).
Then the leading term in the asymptotic expansion of $\tilde{G}_i$ for large $R_P$ and $n$ may
be obtained by the method of stationary phase:  by replacing the phase function $g$ in Eq.\ (25)
with its expanded version (33), approximating the coefficient $f_i$ of the rapidly oscillating exponential with its limiting value
$$f_i\big\vert_O={r_P}^{-1}\big(\sin\xi\quad-1\quad1\big)\eqno(38)$$
at $O$, and extending the range of integration $\Delta\varphi$ to $(\varphi_O-\pi,\varphi_O+\pi)$ (see~\cite{AAS7,AAS72,AAS8}).

The integral that appears in the resulting expression,   
%$$\eqalignno{
\begin{eqnarray*}
{\tilde{G}}_i\sim&(2\pi)^{-1}\exp\{{\rm i}n[{\hat\varphi}-\phi_O-(\eta^2+\zeta^2)/(2{\hat r}_P)]+{\rm i}\Omega\varphi_O/\omega\}\cr
&\times\int_{-\pi}^\pi d\xi\, f_i\big\vert_O \exp\{-{\rm i}[(n-\Omega/\omega)\xi-n{\hat r}\sin\xi]\},\quad&(39)
\end{eqnarray*}
%\cr}$$
can be evaluated in terms of an Anger function and its derivative: 
%$$\eqalignno{
\begin{eqnarray*}
{\tilde{G}}_i\sim&-{r_P}^{-1}\exp\{{\rm i}n[{\hat\varphi}-\phi_O-(\eta^2+\zeta^2)/(2{\hat r}_P)]+{\rm i}\Omega\varphi_O/\omega\}\cr
&\times\big[{\rm i}{{\bf J}^\prime}_{n-\Omega/\omega}(n{\hat r})\quad{\bf J}_{n-\Omega/\omega}(n{\hat r})\quad-{\bf J}_{n-\Omega/\omega}(n{\hat r})\big],\quad&(40)
\end{eqnarray*}
%\cr}$$
where ${\bf J}$ and ${\bf J}^\prime$ are Anger's function and the derivative of Anger's function
with respect to its argument and the symbol $\sim$ denotes asymptotic approximation.
The Anger function ${\bf J}_\nu(\chi)$ is defined by
%$$\eqalignno{
\begin{eqnarray*}
{\bf J}_\nu(\chi)\equiv&(2\pi)^{-1}\int_{-\pi}^\pi d\xi\, \exp[-{\rm i}(\nu\xi-\chi\sin\xi)]\qquad\qquad\qquad\qquad\qquad\qquad\qquad\qquad\cr
&=J_\nu(\chi)+\pi^{-1}\sin(\nu\pi)\int_0^\infty d\tau\exp(-\nu\tau-\chi\sinh\tau),\qquad\qquad\qquad\qquad\qquad&(41)
\end{eqnarray*}
%\cr}$$
in which $J_\nu(\chi)$ is the Bessel function of the first kind~\cite{AAS9,AAS10}.

There is no difference between an Anger and a Bessel function when $\nu$ is an integer.
The second integral in Eq.\ (41), which constitutes the difference between these two functions
when $\sin(\nu\pi)\neq0$, has the asymptotic expansion
%$$\eqalignno{
\begin{eqnarray*}
\int_0^\infty d\tau\exp(-\nu\tau-\chi\sinh\tau)\sim&(1+\nu/\chi)^{-1}\chi^{-1}-(1+\nu/\chi)^{-4}\chi^{-3}\cr
&+(9-\nu/\chi)(1+\nu/\chi)^{-7}\chi^{-5}+\cdots \quad&(42)
\end{eqnarray*}
%\cr}$$
for large $\chi$ and positive $\nu$ (see~\cite{AAS10}).
The leading contributions toward the values of the Anger functions in Eq.\ (40), therefore,
come from the Bessel functions $J_{n-\Omega/\omega}(n{\hat r})$ and ${J^\prime}_{n-\Omega/\omega}(n{\hat r})$
whose asymptotic values for large $n$ decay more slowly than $n^{-1}$ in the superluminal regime.
(Here and in what follows, we treat $\Omega$ as a positive constant.)

When the argument of $J_\nu(\chi)$ is smaller than its order and so can be written
as $\chi=\nu\,{\rm sech}\,\alpha$ for some $\alpha>0$, the asymptotic values,
for large $\nu$, of this Bessel function and its derivative are given by
$$J_\nu(\nu\,{\rm sech}\,\alpha)\sim(2\pi\nu\tanh\alpha)^{-{1\over2}}\exp[\nu(\tanh\alpha-\alpha)],\eqno(43a)$$
and
$${J^\prime}_\nu(\nu\,{\rm sech}\,\alpha)\sim(4\pi\nu/\sinh 2\alpha)^{-{1\over2}}\exp[\nu(\tanh\alpha-\alpha)]\eqno(43b)$$
(see~\cite{AAS9}).  In this regime, the Bessel functions in question
decrease exponentially with increasing $\nu$:  the exponent $\tanh\alpha-\alpha$
is negative for all positive $\alpha$.
But when the argument of $J_\nu(\chi)$ is greater than its order and
can be written as $\chi=\nu\sec\beta$ for some $0<\beta<\pi/2$, we have~\cite{AAS9}
$$J_\nu(\nu\sec\beta)\sim(\textstyle{1\over2}\pi\nu\tan\beta)^{-{1\over2}}\cos[\nu(\tan\beta-\beta)-\textstyle{\pi\over4}],\eqno(44a)$$
and
$${J^\prime}_\nu(\nu\sec\beta)\sim(\pi\nu/\sin 2\beta)^{-{1\over2}}\sin[\nu(\tan\beta-\beta)-\textstyle{\pi\over4}].\eqno(44b)$$
In this case, the Bessel functions in question oscillate with amplitudes that decrease
algebraically, like $\nu^{-{1\over2}}$, with increasing $\nu$.
When $\chi$ and $\nu$ are equal, these functions decay even more slowly:
$J_\nu(\nu)\sim0.44730\times\nu^{-{1\over3}}$ and ${J^\prime}_\nu(\nu)\sim0.41085\times\nu^{-{2\over3}}$ (see~\cite{AAS9}).

The contributions in Eq.\ (24) that arise from the source elements in ${\hat r}<1-\Omega/(n\omega)$,
therefore, are exponentially smaller than those which arise from ${\hat r}>1-\Omega/(n\omega)$:
the asymptotic values of the Green's functions appearing in Eq.\ (24) are proportional either to
$J_{n-\Omega/\omega}(n{\hat r})$ or to ${J^\prime}_{n-\Omega/\omega}(n{\hat r})$ for large $n$.
In particular, the contributions proportional to $J_{n-\Omega/\omega}(n)$ and
${J^\prime}_{n-\Omega/\omega}(n)$, that arise from a source element at the stationary point $O$,
are greater for the terms involving $\Omega$ in Eq.\ (24) than for those involving $-\Omega$.
\subsection{Frequencies of emission from individual volume elements}
\label{VB}
The emission is strongest at the frequency with which the source oscillates in addition to moving,
i.e.\ at a value of $n\omega$ for which the integer $n$ is closest to $\Omega/\omega$.
Figures 6 and 7 respectively show the $n$ dependences of the squares of the
two Bessel functions $J_{n-\Omega/\omega}(n)$ and ${J^\prime}_{n-\Omega/\omega}(n)$ for $\Omega/\omega=15.5$,
each normalized by its value at $n=16$.  The plotted quantities are respectively proportional to
the squares of moduli of the Green's functions ${\tilde G}_2$ and ${\tilde G}_1$ at $O$ [see Eq.\ (40)].
Figure 6 depicts the spectral distribution of the emission that arises from the single comoving point $O$ of
a polarization along the $r$ or the $z$ directions,
and Fig.\ 7 depicts the spectrum of the corresponding emission from a current which flows in the $\varphi$ direction [see Eq.\ (24) for $\theta_P=\pi/2$].

\begin{figure}[tbp]
   \centering
\includegraphics[height=7cm]{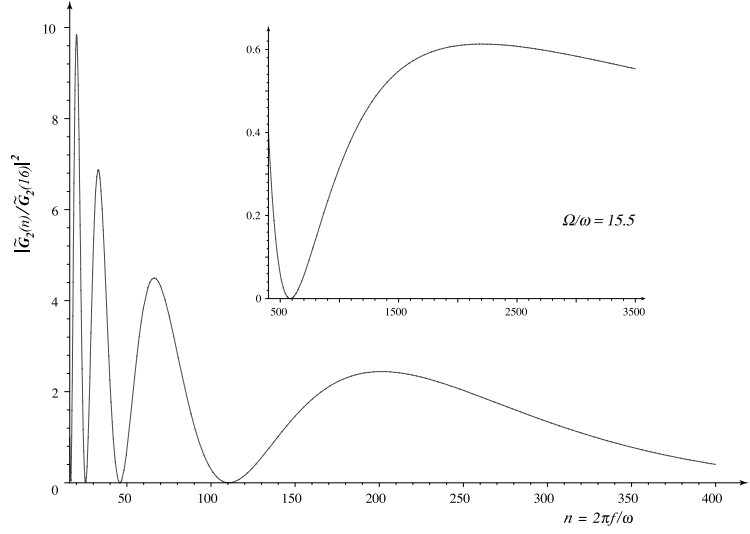}
\caption{Spectral distribution of the Green's function ${\tilde G}_2(n)$
for $\Omega/\omega=15.5$, normalized by the value ${\tilde G}_2(16)$
of this function at a harmonic number ($n=16$) close to $\Omega/\omega$.
(Frequency $f$ and harmonic number $n$ are related via $2\pi f=n\omega$.)
The inset highlights the highest frequency peak of the spectrum.
Note that the ranges of frequencies shown in the figure and its inset are complementary.
This function represents the spectral distribution of the
emission arising from the single comoving point $O$ of
a polarization along the $r$ or the $z$ directions [see Eqs.\ (24) and (40)].}
\end{figure}

\begin{figure}[tbp]
   \centering
\includegraphics[height=7cm]{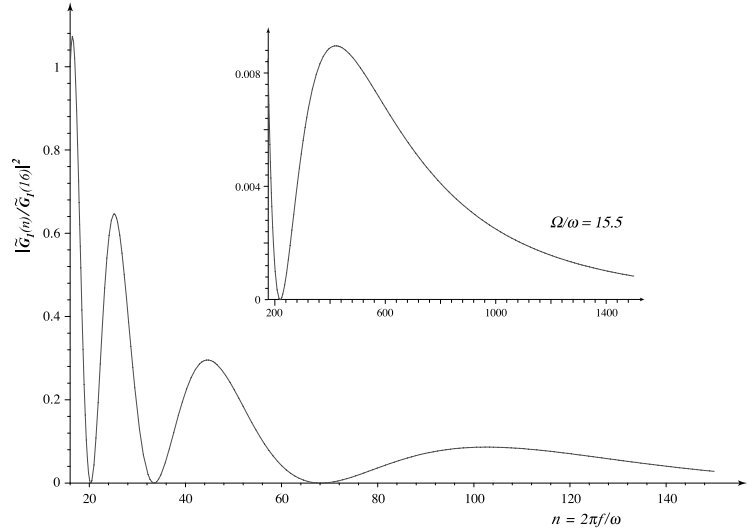}
\caption{Spectral distribution of the Green's function ${\tilde G}_1(n)$,
for $\Omega/\omega=15.5$, normalized by the value ${\tilde G}_1(16)$
of this function at a harmonic number ($n=16$) close to $\Omega/\omega$.
(Frequency $f$ and harmonic number $n$ are related via $2\pi f=n\omega$.)
The inset highlights the highest frequency peak of the spectrum.
Note that the ranges of frequencies shown in the figure and its inset are complementary.
This function represents the spectral distribution of the
emission arising from the single comoving point $O$ of
a polarization current which flows in the $\varphi$ direction
 [see Eq.\ (40) and Eq.\ (24) for $\theta_P=\pi/2$].}
\end{figure}

The spectral distributions shown in Figs.\ 6 and 7 confirm the following result
that was earlier inferred from time-domain considerations in Section~\ref{II}:
the contributions of a rotating point source that is coincident with $O$,
and so approaches the observer with the speed of light and zero acceleration at the retarded time,
are made over a wide range of frequencies and have peaks at harmonic numbers $n\gg1$ which
are of the order of $(\Omega/\omega)^3$ when $\omega\ll\Omega\ll n\omega$.
This is a consequence of the fact that in this regime we have
$J_{n-\Omega/\omega}(n)\sim(2/n)^{1\over3}{\rm Ai}[-(2/n)^{1\over3}\Omega/\omega]$
and ${J^\prime}_{n-\Omega/\omega}(n)\sim-(2/n)^{2\over3}{\rm Ai}^\prime[-(2/n)^{1\over3}\Omega/\omega]$ and
that the Airy function ${\rm Ai}[-(2/n)^{1\over3}\Omega/\omega]$
and its derivative peak where the magnitude of their arguments is of the order of unity~\cite{AAS9}.
\subsection{Superposition of the contributions from the constituent volume elements of the source}
\label{VC}
The Green's function ${\tilde{G}}_i(r,{\hat\varphi},z;n)$ calculated in the preceding
section represents the contribution to $\tilde{\bf E}_n$ of a specific volume element of
the uniformly rotating source:  that which has the azimuthal coordinate ${\hat\varphi}$ at
the time $t=0$ and which moves on a circular orbit of radius $r$ on a plane that is normal
to and crosses the rotation axis at $z$.  To find the radiation
field that arises from an extended source, we must superpose the contributions from the
constituent volume elements of that source, i.e.\ we must insert this Green's function in the integral
representation (24) of the field and perform the integrations with respect to
$r$, ${\hat\varphi}$ and $z$ that extend over the localized region of the rest frame occupied by that source.

Equations (24) and (40) jointly yield
%$$\eqalignno{
\begin{eqnarray*}
{\tilde{\bf E}}_n\sim&\textstyle{1\over2}{{\hat r}_P}^{-1}\exp[-{\rm i}(n\phi_O-\Omega\varphi_O/\omega)]Q_{\hat\varphi}\int_{-\infty}^\infty d{\hat z}\,\exp[-\textstyle{1\over2}{\rm i}n{{\hat r}_P}^{-1}({\hat z}-{\hat z}_P)^2]\cr
&\times\int_0^\infty{\hat r}\, d{\hat r}\,\exp[-\textstyle{1\over2}{\rm i}n{{\hat r}_P}^{-1}({\hat r}-1)^2]{\bf V}+\{m\to-m,\Omega\to -\Omega\},\quad&(45)
\end{eqnarray*}
%\cr}$$
in which
%$$\eqalignno{
\begin{eqnarray*}
Q_{\hat\varphi}\equiv&-\textstyle{1\over2}\sum_{\mu=\mu_\pm}\int_{-\pi}^{+\pi}d{\hat\varphi}\,\mu^2\exp[{\rm i}(n-\mu){\hat\varphi}]\qquad\qquad\qquad\qquad\qquad\qquad\qquad\qquad\qquad\cr
&=(-1)^{n+m}\sin(\pi\Omega/\omega)[{\mu_+}^2(n-\mu_+)^{-1}+{\mu_-}^2(n-\mu_-)^{-1}],\qquad\qquad\qquad&(46)
\end{eqnarray*}
%\cr}$$
and ${\bf V}$ stands for
$${\bf V}\equiv(s_r W_2+s_\varphi W_1){\hat{\bf e}}_\parallel+[\cos\theta_P(s_\varphi W_2-s_r W_1)+\sin\theta_P s_z W_3]{\hat{\bf e}}_\perp\eqno(47)$$
with
$$W_i\equiv\big[{\rm i}{{\bf J}^\prime}_{n-\Omega/\omega}(n{\hat r})\quad{\bf J}_{n-\Omega/\omega}(n{\hat r})\quad-{\bf J}_{n-\Omega/\omega}(n{\hat r})\big].\eqno(48)$$
Here we have replaced $r$, $r_P$ and $z$ by their values in units of the light-cylinder radius
(${\hat r}$, ${\hat r}_P$ and ${\hat z}$) and have used the definition
$\mu_\pm\equiv(\Omega/\omega)\pm m$ to rewrite $\sin[\pi(n-\mu_\pm)]$ as $(-1)^{n+m+1}\sin(\pi\Omega/\omega)$.
The Anger function ${\bf J}$ and its derivative ${\bf J}^\prime$ are defined in Eq.\ (41).

Note that if $\Omega/\omega$ is an integer, the factors $\sin[\pi(n-\mu_\pm)]/(n-\mu_\pm)$ in Eq.\ (46)
would have the value $\pi$ at $n=\mu_\pm$ and would vanish at all other $n$.
When $\Omega$ and $\omega$ are not commensurable, on the other hand, neither the numerators nor the
denominators in these factors can vanish at any $n$.  For non-integral values of $n$, the quantity $Q_{\hat\varphi}$,
and hence the spectrum of the source density (15), is non-zero for all frequencies (see also Section~\ref{II} and Fig.\ 5).
We shall here assume not only that the parameter $\Omega/\omega$ is different from an integer
but also that it is appreciably greater than unity.
Our interest lies primarily in the high-frequency regime where the radiation frequency
$n\omega$ is appreciably greater than the frequencies $\vert m\omega\pm\Omega\vert$
that enter the creation of the source (cf.\ Appendix A).

The leading terms in the asymptotic expansions for large $n$ of the integrals
over ${\hat r}$ and ${\hat z}$ in the expression for ${\tilde{\bf E}}_n$ can both be found
by an elementary version of the method of stationary phase~\cite{AAS7,AAS72,AAS8}.
If the lower and upper limits of the radial interval in which the source densities $s_{r,\varphi,z}$
are non-zero are denoted by ${\hat r}_<<1$ and ${\hat r}_>>1$, respectively,
then the asymptotic value of the integral over ${\hat r}$ in Eq.\ (45) is given by
$$\int_0^\infty{\hat r}\, d{\hat r}\,\exp[-\textstyle{1\over2}{\rm i}n{{\hat r}_P}^{-1}({\hat r}-1)^2]{\bf V}\sim{\bf V}\big\vert_{{\hat r}=1}Q_r,\eqno(49)$$
in which
%$$\eqalignno{
\begin{eqnarray*}
Q_r\equiv&\int_{{\hat r}_<}^{{\hat r}_>}\, d{\hat r}\,\exp[-\textstyle{1\over2}{\rm i}n{{\hat r}_P}^{-1}({\hat r}-1)^2]\qquad\qquad\qquad\qquad\qquad\qquad\qquad\qquad\cr
&=(\pi{\hat r}_P/n)^{1\over2}\{C(\eta_>)-C(\eta_<)-{\rm i}[S(\eta_>)-S(\eta_<)]\},\qquad\qquad\qquad&(50a)
\end{eqnarray*}
%\cr}$$
with
$$\eta_>\equiv[n/(\pi{\hat r}_P)]^{1\over2}({\hat r}_>-1),\eqno(50b)$$
$$\eta_<\equiv[n/(\pi{\hat r}_P)]^{1\over2}({\hat r}_<-1),\eqno(50c)$$
and the functions $C(\eta)$ and $S(\eta)$ are the Fresnel integrals~\cite{AAS9}.
We have integrated over the interval $({\hat r}_<,{\hat r}_>)$, rather than $(-\infty,\infty)$,
in order to obtain expressions that are valid also within the Fresnel
distance $r_P\sim(r_>-r_<)^2(n\omega)/(\pi c)$ from the source.
For an observation point that lies at infinity, the integration with respect to
${\hat r}$ may be directly performed over $(-\infty,\infty)$, without introducing ${\hat r}_<$ and ${\hat r}_>$.

If we now insert Eq.\ (49) in Eq.\ (45), and carry out the remaining integration
with respect to ${\hat z}$ in a similar way, we arrive at
$${\tilde{\bf E}}_n\sim \textstyle{1\over2}{{\hat r}_P}^{-1}\exp[-{\rm i}(n\phi_O-\Omega\varphi_O/\omega)]Q_{\hat\varphi}Q_rQ_z{\bf V}\big\vert_{{\hat r}=1,{\hat z}={\hat z}_P}+\{m\to-m,\Omega\to -\Omega\},\eqno(51)$$
in which
%$$\eqalignno{
\begin{eqnarray*}
Q_z\equiv&\int_{{\hat z}_<}^{{\hat z}_>}\, d{\hat z}\,\exp[-\textstyle{1\over2}{\rm i}n{{\hat r}_P}^{-1}({\hat z}-{\hat z}_P)^2]\qquad\qquad\qquad\qquad\qquad\qquad\qquad\qquad\cr
&=(\pi{\hat r}_P/n)^{1\over2}\{C(\zeta_>)-C(\zeta_<)-{\rm i}[S(\zeta_>)-S(\zeta_<)]\},\qquad\qquad\qquad&(52a)
\end{eqnarray*}
%\cr}$$
with
$$\zeta_>\equiv[n/(\pi{\hat r}_P)]^{1\over2}({\hat z}_>-{\hat z}_P),\eqno(52b)$$
and
$$\zeta_<\equiv[n/(\pi{\hat r}_P)]^{1\over2}({\hat z}_<-{\hat z}_P).\eqno(52c)$$
Here, $z_>>z_P$ and $z_<<z_P$ are the lower and upper limits on the extent of the
source distribution in the $z$ direction.
(The coordinates $r$ and $z$, without the superscript ${\hat{}}\,$, are measured in standard units
of length rather than in units of the light-cylinder radius $c/\omega$.)

When the $z$ interval $z_>-z_<$ that is occupied by the source is appreciably smaller than the radius
of the light cylinder, the observer would have to have a colatitude $\theta_P$ that is
quite close to $\pi/2$ for the stationary point $O$ to lie within the source distribution.
From Eqs.\ (47) and (48) for $\theta_P=\pi/2$, it therefore follows that
$${\bf V}\big\vert_{{\hat r}=1,{\hat z}={\hat z}_P}={\bf J}_{n-\Omega/\omega}(n)(s_{rO}{\hat{\bf e}}_\parallel-s_{zO}{\hat{\bf e}}_\perp)+{\rm i}{{\bf J}^\prime}_{n-\Omega/\omega}(n)s_{\varphi O}{\hat{\bf e}}_\parallel,\eqno(53)$$
in which $s_{rO,\varphi O,zO}$ stand for the values of $s_{r,\varphi,z}$ at the
stationary point $O$.
Thus the contributions of the poloidal components of the
polarization ($s_r$ and $s_z$) to the radiation field are both stronger than, and $90$ degrees out of
phase with, the contribution of the toriodal component $s_\varphi$ (see Table~2).

\subsection{The radiated power and its spectral distribution}
\label{VD}
The expression that is given by Eq.\ (51) for the Fourier component $\tilde{\bf E}_n$
of the radiation field in the plane of rotation applies to any frequency $n\omega$ for which $n\gg1$,
provided of course that the distance $r_P$ of the observer appreciably exceeds both the
radius $c/\omega$ of the light cylinder and the dimensions $r_>-r_<$ and $z_>-z_<$ of the source.
If $n\omega$ is also much greater than the frequencies
$\vert m\omega\pm\Omega\vert$ that enter the creation of the source,
the asymptotic value of the radiation field for large frequency reduces to
%$$\eqalignno{
\begin{eqnarray*}
\tilde{\bf E}_n\sim&(-1)^{n+m}{{\hat r}_P}^{-1}\exp[-{\rm i}(n\phi_O-\Omega\varphi_O/\omega)](m^2+\Omega^2/\omega^2)\sin(\pi\Omega/\omega)Q_rQ_z\cr
&\times n^{-1}\big[J_{n-\Omega/\omega}(n)(s_{rO}{\hat{\bf e}}_\parallel-s_{zO}{\hat{\bf e}}_\perp)+{\rm i}{J^\prime}_{n-\Omega/\omega}(n)s_{\varphi O}{\hat{\bf e}}_\parallel\big],\qquad&(54)
%\cr}$$
\end{eqnarray*}
for both the additional terms in the Anger functions [Eq.\ (41)]
and the terms associated with $-\Omega$ [Eq.\ 51)] would be negligibly small.
This expression holds not only in the far zone but also within the Fresnel zone.
The Fresnel distance from the source, which designates the boundary between the near
and far zones at a given frequency $n\omega$, is defined by $r_P\simeq[n\omega/(\pi c)]\,{\rm max}\,\{(r_>-r_<)^2 , (z_>-z_<)^2\}$ in the present case.

When the observation point lies closer to the source than the Fresnel distance,
the arguments of the Fresnel integrals $C$ and $S$ in Eqs.\ (50) and (52) are large and so $Q_r$ and $Q_z$ assume the values
$$Q_r\sim(2\pi{\hat r}_P/n)^{1\over2}\exp(-{\rm i}\pi/4),\qquad{\hat r}_P\ll(n/\pi)({\hat r}_>-{\hat r}_<)^2,\eqno(55a)$$
$$Q_z\sim(2\pi{\hat r}_P/n)^{1\over2}\exp(-{\rm i}\pi/4),\qquad{\hat r}_P\ll(n/\pi)({\hat z}_>-{\hat z}_<)^2.\eqno(55b)$$
Beyond the Fresnel distance from the source, on the other hand, the limiting values of $C$ and $S$ are such~\cite{AAS9} that
$$Q_r\sim{\hat r}_>-{\hat r}_<,\qquad{\hat r}_P\gg(n/\pi)({\hat r}_>-{\hat r}_<)^2,\eqno(56a)$$
$$Q_z\sim{\hat z}_>-{\hat z}_<,\qquad{\hat r}_P\gg(n/\pi)({\hat z}_>-{\hat z}_<)^2\eqno(56b)$$
for in the far zone the arguments of the oscillating exponentials in the integrands in Eqs.\ (50a) and (52a) tend to zero.
The inclusion of higher order terms in the expansion of $g$ in powers of ${\hat r}-1$ and ${\hat z}-{\hat z}_P$
would not alter this result:  the coefficients of these higher order terms contain correspondingly higher powers of ${{\hat r}_P}^{-1}$.

The radiated power received by an observer in the far zone per harmonic per unit solid angle is given by
$dP_n/d\Omega_P=c{R_P}^2\vert{\tilde{\bf E}}_n\vert^2/(2\pi)$, where $d\Omega_P$
denotes the element $\sin\theta_P\,d\theta_P\,d\varphi_P$ of solid angle in the space of observation points.
According to Eqs.\ (54) and (56), this power has the following asymptotic value for high frequency and $\theta_P=\pi/2$:
%$$\eqalignno{
\begin{eqnarray*}
dP_n/d\Omega_P\sim&(2\pi c)^{-1}\omega^2(m^2+\Omega^2/\omega^2)^2\sin^2(\pi\Omega/\omega)(r_>-r_<)^2(z_>-z_<)^2\cr
&\times n^{-2}\big\vert J_{n-\Omega/\omega}(n)(s_{rO}{\hat{\bf e}}_\parallel-s_{zO}{\hat{\bf e}}_\perp)+{\rm i}{J^\prime}_{n-\Omega/\omega}(n)s_{\varphi O}{\hat{\bf e}}_\parallel\big\vert^2.\qquad&(57)
\end{eqnarray*}
%\cr}$$
It depends on the harmonic number $n$ like $n^{-2}{J^2}_{n-\Omega/\omega}(n)$ if the
polarization lies in the $z$ or $r$ directions, and like $n^{-2}{{J^\prime}^2}_{n-\Omega/\omega}(n)$
if the polarization lies in the $\varphi$ direction.

Figures 8 and 9 show the emitted power by a poloidal polarization current
($s_r$ or $s_z$) for $\Omega/\omega=15.5$ and $m=72$.
The amplitude of this power has in both figures been normalized by its value at the harmonic number
closest to the source frequency $m\omega+\Omega$, i.e.\ by the power that is emitted into $n=87$.
[Note that the peak emission of the device described in Appendix A is always close to $f=(m \omega + \Omega)/2\pi$;
in the case of $\Omega/\omega=$ integer, the {\it only} emission occurs at this frequency
and its companion $f=(m \omega - \Omega)/2\pi$.
We show elsewhere~\cite{newpaper} that, irrespective of the
value of $\Omega/\omega$, the non-spherically-decaying
part of the emission only contains the frequencies $f=\vert m \omega \pm \Omega\vert/2\pi$.]

The curve shown in Figs.\ 8 and 9 decreases monotonically from one (the maximum power)
at $n=87$ to zero at its minimum (see Fig.\ 9).
As a result of the presence of the extra factor ${Q_{\hat\varphi}}^2\sim n^{-2}$,
the value of $n$ at which the curve in Fig.\ 8 attains its high-frequency maximum is
somewhat lower than that of the corresponding curve shown in the inset to Fig.\ 6.
The order of magnitude of this $n$, however, is still given by $(\Omega/\omega)^3$:
the result earlier encountered in Sections~\ref{II} and \ref{VA}
in the context of a point source thus applies also to an extended source.

\begin{figure}[tbp]
   \centering
\includegraphics[height=7cm]{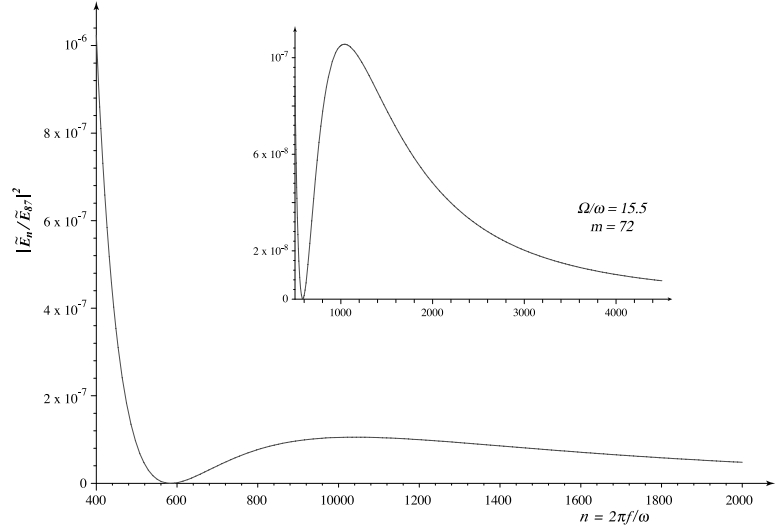}
\caption{Spectral distribution of the radiation that is generated by a poloidal
polarization current for $\Omega/\omega=15.5$ and $m=72$.
The normalization factor is the value ${\tilde{\bf E}}_{87}$ of the radiation field
at a harmonic number ($n=87$) close to the peak emission at
$n=(m+\Omega/\omega)$ or $f=(m\omega+\Omega)/2\pi$ (see Fig.~9);
frequency $f$ and harmonic number $n$ are related via $2\pi f=n\omega$.
The plotted quantity has the value $1$ at $n=87$ and decreases monotonically over the range
$87<n<400$ of harmonic numbers not shown here (see Fig.\ 9).  The inset
highlights the highest frequency peak of the spectrum.}
\end{figure}

\begin{figure}[tbp]
   \centering
\includegraphics[height=7cm]{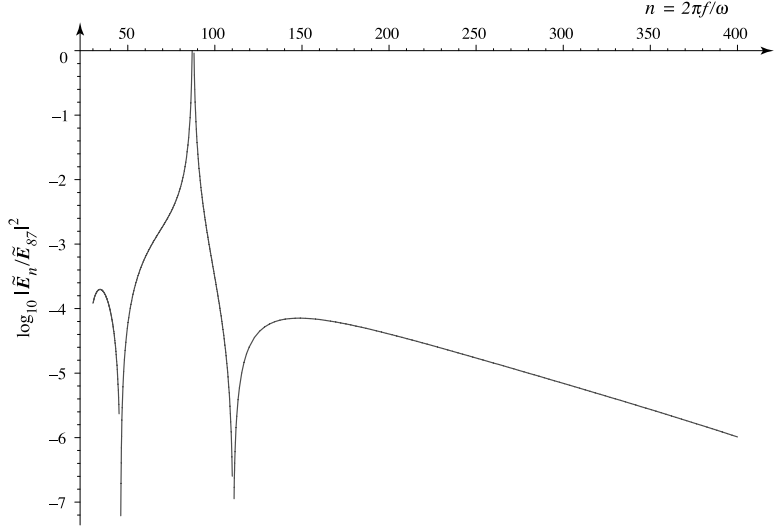}
\caption{Logarithm of the normalized intensity, $\log_{10}\vert{\tilde{\bf E}}_n/{\tilde{\bf E}}_{87}\vert^2$,
shown in Fig.\ 8 over the lower range $30<n<400$ of harmonic numbers (recall that
$\Omega/\omega=15.5$ and $m=72$).  Note that the ranges of frequencies in this figure and in Fig.\ 8 are complementary.  The contribution from $\vert\mu_-\omega\vert=m\omega-\Omega$ (see Table~1) to the radiation field is in this case too small to make a difference to the figure even at these lower frequencies.}
\end{figure}

Note that, according to Eqs.\ (54) and (55), the amplitude of $\tilde{\bf E}_n$ is independent of the distance
$r_P$ from the source throughout the Fresnel zone:
it diminishes like ${r_P}^{-1}$ only beyond the Fresnel distance,
i.e.\ for $r_P\gg[n\omega/(\pi c)]\,{\rm max}\,\{(r_>-r_<)^2 , (z_>-z_<)^2\}$.
This non-spherical decay of the field amplitude within the Fresnel zone is encountered whenever
the radiated wave fronts have envelopes; the field of \v Cerenkov radiation,
for instance, decays like ${R_P}^{-{1\over2}}$ within its Fresnel zone.
However, the constancy of the amplitude of the radiation with distance is here compensated by a
steeper dependence of this amplitude on frequency.
At the position of an observer who is closer to the source than the Fresnel distance,
the rate of decay of $dP_n/d\Omega_P$ with $n$ is by the factor $n^{-2}$
higher than that of the power which is radiated in the far zone.
\section{The radiation field outside the plane of the source's orbit}
\label{VI}
\subsection{Asymptotic expansion of the Green's function}
\label{VIA}
When the observation point is located at a colatitude $\theta_P$ which
is different from $\pi/2$, the leading contribution to the asymptotic value of the
radiation field for large ${\hat R}_P$ and large $n$ comes from those volume elements of
the source that lie on the curve described in Eq.\ (28), the curve we designated as $C$.
For ${\hat R}_P\gg1$, $C$ is coincident with the cusp curve of the bifurcation surface, $C_b$,
and the source points in question are those which approach the observer with the speed of light
and zero acceleration at the retarded time (see Fig.\ 2).  The integrations with respect to
$\varphi$ and $r$ in Eqs.\ (24) and (25) can, as a result, be evaluated by the method of stationary phase once again.

It can be seen from the far-field limit of Eq.\ (29) that the cusp curve $C_b$ would intersect the source
distribution if $\theta_P$ lies in the interval $\vert\theta_P-{\pi\over2}\vert\leq\arccos\, (1/{{\hat r}_>})$,
where $r_>$ is the radial coordinate of the outer boundary of the source~\cite{AAS1}.
Hence, the range of values of $\theta_P$ to which the following analysis is
applicable would be as wide as $(0,\pi)$ if the extent $r_>-c/\omega$ of
the superlumially moving part of the source is comparable to the radius $c/\omega$ of the light cylinder.

For $\theta_P\neq\pi/2$, the relevant expansion of the phase function $g$ about the locus of
its stationary points is that found in Eq.\ (37).  The counterpart of Eq.\ (39) is therefore given by
%$$\eqalignno{
\begin{eqnarray*}
{\tilde{G}}_i\sim&(2\pi)^{-1}\exp\{{\rm i}[n({\hat\varphi}-\phi_C-\textstyle{1\over2}{{\hat R}_C}^{-1}\rho^2)+\Omega\varphi_C/\omega]\}\qquad\qquad\cr
&\times\int_{-\pi}^{\pi}d\mu\, f_i\big\vert_C\exp\{-{\rm i}[(n-\Omega/\omega)\mu-n({\hat r}/{\hat r}_C)\sin\mu]\}\qquad\qquad&(58)
\end{eqnarray*}
%\cr}$$
in which ${\hat r}_C$, $\varphi_C$ and ${\hat R}_C$, defined in Eqs.\ (28) and (36), have their far-field values
$${\hat r}_C\simeq\csc\theta_P,\quad\varphi_C\simeq\varphi_P+3\pi/2,\quad{\rm and}\quad{\hat R}_C\simeq{\hat R}_P-{\hat z}\cos\theta_P,\eqno(59)$$
and $f_i\big\vert_C$ its limiting value
$$f_i\big\vert_C={R_P}^{-1}\big(\sin\mu\quad-1\quad1\big).\eqno(60)$$
[The term ${{\hat R}_C}^{-1}\mu^2$ in Eq.\ (37) has, for the reasons given in Section~\ref{IV}, been omitted here.]

The integral that appears in Eq.\ (58) is expressible in terms of an Anger function and its derivative:
$${\tilde{G}}_i\sim-{R_P}^{-1}\exp\{{\rm i}[n({\hat\varphi}-\textstyle{3\over2}\pi-{\hat R}_P+{\hat z}\cos\theta_P-\textstyle{1\over2}{{\hat R}_P}^{-1}\rho^2)+\Omega\varphi_C/\omega]\}{\bar W}_i\eqno(61)$$
with
$${\bar W}_i\equiv\big[{\rm i}{{\bf J}^\prime}_{n-\Omega/\omega}(n{\hat r}\sin\theta_P)\quad{\bf J}_{n-\Omega/\omega}(n{\hat r}\sin\theta_P)\quad-{\bf J}_{n-\Omega/\omega}(n{\hat r}\sin\theta_P)\big].\eqno(62)$$
Hence, Eqs.\ (24) and (61) jointly yield
%$$\eqalignno{
\begin{eqnarray*}
{\tilde{\bf E}}_n\sim&\textstyle{1\over2}{{\hat R}_P}^{-1}\exp\{-{\rm i}[n({\hat R}_P+\textstyle{3\over2}\pi)-\Omega\varphi_C/\omega]\}Q_{\hat\varphi}\int_{-\infty}^\infty d{\hat z}\,\exp({\rm i}n{\hat z}\cos\theta_P)\qquad\qquad\cr
&\times\int_0^\infty{\hat r}\, d{\hat r}\,\exp[-\textstyle{1\over2}{\rm i}n{{\hat R}_P}^{-1}({\hat r}-\csc\theta_P)^2]{\bar{\bf V}}+\{m\to-m,\Omega\to -\Omega\},\qquad\qquad&(63)
%\cr}$$
\end{eqnarray*}
where ${\bar{\bf V}}$ differs from the vector ${\bf V}$ defined in Eq.\ (47)
only in that $W_i$ is everywhere replaced in it by ${\bar W}_i$ [cf.\ Eq.\ (62)].
The quantity $Q_{\hat\varphi}$, resulting from the integration with respect to ${\hat\varphi}$, is the same here as in Eq.\ (46).

The dominant contribution towards the asymptotic approximation to the integral over ${\hat r}$ for
high frequency arises from the value of its intgrand at the stationary point
${\hat r}=\csc\theta_P$ of its phase, just as in Eq.\ (49).
The ${\hat r}$ integral in Eq.\ (63) therefore has the asymptotic value
$$\int_0^\infty{\hat r}\, d{\hat r}\,\exp[-\textstyle{1\over2}{\rm i}n{{\hat R}_P}^{-1}({\hat r}-\csc\theta_P)^2]{\bar{\bf V}}\sim\csc\theta_P{\bar{\bf V}}\big\vert_{{\hat r}=\csc\theta_P}{\bar Q}_r,\eqno(64)$$
where
$${\bar Q}_r=(\pi{\hat R}_P/n)^{1\over2}\{C({\bar\eta}_>)-C({\bar\eta}_<)-{\rm i}[S({\bar\eta}_>)-S({\bar\eta}_<)]\},\eqno(65a)$$
with
$${\bar\eta}_>\equiv[n/(\pi{\hat R}_P)]^{1\over2}({\hat r}_>-\csc\theta_P),\eqno(65b)$$
$${\bar\eta}_<\equiv[n/(\pi{\hat R}_P)]^{1\over2}({\hat r}_<-\csc\theta_P),\eqno(65c)$$
and ${\hat r}_<<\csc\theta_P$ and ${\hat r}_>>\csc\theta_P$ are the lower and upper
limits of the radial interval in which the source densities $s_{r,\varphi,z}$ are non-zero.
\subsection{The radiated power and its spectral distribution}
\label{VIB}
From Eqs.\ (63) and (64), it now follows that
$${\tilde{\bf E}}_n\sim\textstyle{1\over2}{{\hat r}_P}^{-1}\exp\{-{\rm i}[n({\hat R}_P+\textstyle{3\over2}\pi)-\Omega\varphi_C/\omega]\}Q_{\hat\varphi}{\bar Q}_r{\bar{\bf Q}}_z+\{m\to-m,\Omega\to -\Omega\},\eqno(66)$$
in which
%$$\eqalignno{
\begin{eqnarray*}
{\bar{\bf Q}}_z\equiv&\big[{\bar s}_r{\bf J}_{n-\Omega/\omega}(n)+{\rm i}{\bar s}_\varphi{{\bf J}^\prime}_{n-\Omega/\omega}(n)\big]{\hat{\bf e}}_\parallel+\big[({\bar s}_\varphi\cos\theta_P\qquad\qquad\qquad\cr
&-{\bar s}_z\sin\theta_P){\bf J}_{n-\Omega/\omega}(n)-{\rm i}{\bar s}_r\cos\theta_P{{\bf J}^\prime}_{n-\Omega/\omega}(n)\big]{\hat{\bf e}}_\perp,\qquad\qquad&(67)
\end{eqnarray*}
%}$$
with
$${\bar s}_{r,\varphi,z}\equiv\int_{-\infty}^\infty d{\hat z}\,\exp({\rm i}n{\hat z}\cos\theta_P)s_{r,\varphi,z}\big\vert_C.\eqno(68)$$
The source densities $s_{r,\varphi,z}$ in ${\bar s}_{r,\varphi,z}$ are evaluated along
curve $C$ and so are functions of $z$ only.
Note that the Anger functions that appear in this expression are precisely the same as
those appearing in Eq.\ (53) and depicted in Figs.\ 6 and 7:
the field ${\tilde{\bf E}}_n$ depends on $\theta_P$ only through the other factors in Eq.\ (67).

The corresponding expression for the radiated power per harmonic per unit solid angle is therefore given by
$$dP_n/d\Omega_P\sim(2\pi)^{-1}c\csc^2\theta_P\sin^2(\pi\Omega/\omega)(m^2+\Omega^2/\omega^2)^2(r_>-r_<)^2n^{-2}\vert {\bar{\bf Q}}_z\vert^2\eqno(69)$$
when the observation point lies beyond the Fresnel distance
$R_P\sim(n\omega/c)(r_>-r_<)^2$ and $n$ appreciably exceeds $\vert m\pm\Omega/\omega\vert$.
Here, as in Eq.\ (54), both the additional terms in the Anger functions [Eq.\ (41)]
and the terms associated with $-\Omega$ [Eq.\ (66)] are negligibly small,
so that the Anger functions in $\vert{\bar{\bf Q}}_z\vert$ may be approximated by Bessel functions.

The rate of decay of $\vert{\bar{\bf Q}}_z\vert$ with the harmonic number $n$ depends on the
length scale of variations of $s_{r,\varphi,z}\vert_C$ with $z$.
The smoother the distribution of these source densities in $z$, the faster would be the rate
of decay of $\vert{\bar{\bf Q}}_z\vert$ with $n$.
In the case of a source with a limited extent in $z$ whose density falls to zero sharply at its $z$ boundaries,
such as the source described in Appendix A, the quantity $\vert{\bar{\bf Q}}_z\vert$ decays like $n^{-1}$ for large $n$.
The dependence of the poloidal part of the above $dP_n/d\Omega_P$ on $n$ for $\vert{\bar{\bf Q}}_z\vert\sim n^{-1}$
differs from the corresponding spectrum shown in Fig.\ 8 only in that its amplitude is reduced and its peak is shifted to a slightly lower value of $n$.

In the high-frequency regime [where $J_{n-\Omega/\omega}(n)$
decreases like $n^{-{1\over3}}$], the dominant terms of $dP_n/d\Omega_P$ in
Eqs.\ (57) and (69) decay according to a power law $n^{-\alpha}$:  the index $\alpha$ in this power
law has the value $8/3$ in the plane of source's orbit and a value $\ge11/3$ outside that plane.
\subsection{Polarization state of the emitted radiation}
Having examined the dependence of the emitted radiation on the polar angle $\theta_P$, we finally summarise its polarization and indicate
how this is related to the direction {\bf s} of the
emitting current within the device described in Appendix A. This summary, for cases in which one of the cylindrical components of ${\bf s}$ is appreciably larger than the other components, is
shown in Table~2.

\begin{table}[tbp] \centering
\begin{tabular}{|l|l|l|}
\hline
 ~ & $\theta_P={\pi\over2}$ & $\theta_P\neq{\pi\over2}$ \\ \hline
$s_r\neq0,s_\varphi=s_z=0$ & linear, ${\hat{\bf e}}_\parallel$, phase$=0$ & elliptic \\ \hline
$s_\varphi\neq0,s_r=s_z=0$ & linear, ${\hat{\bf e}}_\parallel$, phase$={\pi\over2}$ & elliptic \\ \hline
$s_z\neq0,s_r=s_\varphi=0$ & linear, ${\hat{\bf e}}_\perp$, phase$=\pi$ & linear, ${\hat{\bf e}}_\perp$ \\ \hline
\end{tabular}
\caption{The state of polarization of the radiation in and out of the plane of rotation for different
orientations of the emitting polarization current}
\label{table2}
\end{table}

\section{Comparison with \v Cerenkov, synchrotron and dipole radiations}
\label{VII}
The features that the above analysis has in common with that of \v Cerenkov radiation,
such as the multi-valuedness of the retarded time and the presence of an
envelope of wave fronts at which the phases of the radiation integrals are stationary, are easily recognizable.
To compare our results with those that are familiar from the analyses of synchrotron and
dipole radiations, however, we need to give a brief parallel account of certain features of these
more conventional emission processes in the present notation.

The electric current density for a uniformly rotating point source (i.e.\ the source of synchrotron radiation) is described by    
$${\bf j}(r,\varphi,z,t)=q\,\omega\delta(r-r_0)\delta({\hat\varphi}-{\hat\varphi}_0)\delta(z-z_0){\hat {\bf e}}_\varphi,\eqno(70)$$ 
in which the charge $q$ and the coordinates $(r_0,{\hat\varphi}_0,z_0)$ are all constant.  (Recall that ${\hat\varphi}\equiv\varphi-\omega t$.)\par  
Insertion of this source density in Eq.\ (14) results in
%$$\eqalignno{
\begin{eqnarray*}
{\bf E}\simeq& q(\omega/c)^2\int_0^\infty r\, dr\int_{-\infty}^{+\infty}dz\int_{-\pi}^{+\pi}d{\hat\varphi}\,\delta(r-r_0)\delta^\prime({\hat\varphi}-{\hat\varphi}_0)\delta(z-z_0)\cr
&\times(G_1{\hat{\bf e}}_\parallel+\cos\theta_PG_2{\hat{\bf e}}_\perp),\qquad\qquad&(71)
%\cr}$$
\end{eqnarray*}
where $G_1$ and $G_2$ are defined by the same integral as that in Eq.\ (20) with $\Omega=0$.
The time dependence of the integrand in Eq.\ (71) can once again be expanded into a Fourier series,
not because ${\hat\varphi}$ is limited to an interval of length $2\pi$ (here ${\hat\varphi}$ has a single
fixed value, ${\hat\varphi}_0$) but because the source density is periodic.
Evaluation of the trivial integrals with respect to $(r,{\hat\varphi},z)$ thus results in the following counterpart of Eq.\ (24):
$${\tilde{\bf E}}_n\simeq-{\rm i}n(\omega/c)^2q\, r_0({\tilde G}_1{\hat{\bf e}}_\parallel+\cos\theta_P{\tilde G}_2{\hat{\bf e}}_\perp).\eqno(72)$$
The definitions of ${\tilde G}_1$ and ${\tilde G}_2$ in this expression differ
from those appearing in Eqs.\ (25) and (26) only in that in them
$\Omega=0$, $\Delta\varphi$ is $(\varphi_P-\pi,\varphi_P+\pi)$, and $(r,{\hat\varphi},z)$
are everywhere replaced by $(r_0,{\hat\varphi}_0,z_0)$.

Once the phase function $g$ in these definitions is approximated by its far-field value [Eq.\ (30)]
and the integrations with respect to $\varphi$ are performed, we arrive at expressions for ${\tilde G}_i$
whose limiting values for ${\hat r}_0\to{\hat r}_C$ are identical to the expressions that
follow from the far-field versions of Eqs.\ (61) and (62) when $\Omega=0$.
 Insertion of the resulting expressions for ${\tilde G}_1$ and ${\tilde G}_2$ in Eq.\ (72) then leads to
%$$\eqalignno{
\begin{eqnarray*}
{\tilde{\bf E}}_n\sim&{\rm i}n\,q(\omega/c)^2(r_0/R_P)\exp[{\rm i}n({\hat\varphi}_0-\textstyle{3\over2}\pi-{\hat R}_P+{\hat z}_0\cos\theta_P)]\cr
&\times[{\rm i}{J^\prime}_n(n{\hat r}_0\sin\theta_P){\hat{\bf e}}_\parallel+{{\hat r}_0}^{-1}J_n(n{\hat r}_0\sin\theta_P)\cot\theta_P{\hat{\bf e}}_\perp)],&(73)
%\cr}$$
\end{eqnarray*}
i.e.\ to the familiar field of synchrotron radiation (cf.~\cite{AAS11}).

The Bessel functions that appear in Eq.\ (73) have arguments that are smaller than their orders [as in Eq.\ (43)]
and so decrease exponentially with increasing $n$:  the speed $r_0\omega$ of
the source is subluminal (${\hat r}_0\equiv r_0\omega/c<1$) in the synchrotron process.
In contrast, the Bessel functions that appear in the superluminal regime have arguments
which could equal or exceed their orders and so oscillate with an amplitude that decreases
algebraically, like $n^{-{1\over3}}$, $n^{-{1\over2}}$, or $n^{-{2\over3}}$ [see Eq.\ (44) and the paragraph following it].
Were it to exist, a superluminally rotating {\it point} source would therefore
be a much more efficient source of high-frequency radiation than a subluminally rotating one.
In fact, as can be more directly seen from an analysis in the time domain~\cite{AAS1}, the field of a
(hypothetical) superluminally moving point source is infinitely strong, i.e.\ has a divergent value,
on the envelope of the wave fronts that emanate from it.

On the other hand, by virtue of being point-like, the source of synchrotron radiation has a
spectrum that already contains all frequencies.
Because the spectral distribution of the source in Eq.\ (70) is independent of frequency,
the spectrum of synchrotron radiation is determined solely by the spectral distribution of its Green's function.
An extended source is radically different in this respect:
the spectral content of an extended source of the same type, i.e.\ a rotating source with a
density ${\bf j}={\bf j}(r,{\hat\varphi}, z)$ whose strength is time
independent in its own rest frame, is limited to only those wavelengths
which characterize the length scales of its variations in ${\hat\varphi}$.
Had $\Omega$ been zero for the source described in Eq.\ (7),
the spectrum of this source, and hence that of the radiation that arose from it,
would have contained only the single frequency $m\omega$.
[What endows the volume source (7) with the broad---albeit rapidly decaying---spectral
distribution given in Eq.\ (9) is a new effect, having to do with centripetal acceleration,
which would not come into play unless the source strength varies with time.]

In the subluminal regime, volume-distributed charges and currents are typically weak
as sources of radiation:  the contributions from their separate volume elements
(those more distant from each other than one radiation wavelength)
would arrive at the observer with differing phases and so would, as a rule, superpose incoherently.
The contributions to the radiation field that arise from the source elements
located in the vicinity of a stationary point of the optical distance $g$,
however, are an exception to this rule.
In the case of a superluminally moving extended source,
where the derivatives of $g$ have zeros, it is possible for the contributions
from source elements that are more distant than a wavelength to superpose coherently.  Not only does the radiation described in Section~\ref{V}
receive contibutions from an interval of retarded time that is by a factor
of the order of $n^{2\over3}$ longer than its period $2\pi/(n\omega)$,
but also the source elements that contribute coherently towards this
emission occupy $r$ and $z$ intervals that are by the factors $\sim n^{1\over2}$
and $\sim n$ longer than the wavelength $2\pi c/(n\omega)$ in the Fresnel
and the far zones, respectively [see the values of $Q_r$ and $Q_z$ found in Eqs.\ (55) and (56)].

To compare the present results with those that are familiar from the analysis of
dipole radiation, let us consider a case in which the polarization ${\bf P}$ lies in the
$z$ direction (i.e.\ $s_r=s_\varphi=0$) and  $\Omega/\omega$ is an integer.
The only non-zero values of the quantity $Q_{\hat\varphi}$ appearing in Eq.\ (46)
are in this case those which occur when $n=(\Omega/\omega)\pm m$,
so that the amplitude of the radiation field ${\tilde{\bf E}}_n$ reduces to
$$\vert{\tilde{\bf E}}_n\vert\sim\pi{{\hat r}_P}^{-1}\vert{\bf s}\vert({\hat r}_>-{\hat r}_<)({\hat z}_>-{\hat z}_<)n^2\vert J_{n-\Omega/\omega}(n)\vert\eqno(74)$$
for $n=m+\Omega/\omega$ [see Eqs.\ (45), (54) and (56)].\par
The ${\hat\varphi}$ extent of the contributing part of the source is of
the order of a wavelength, $2\pi c/(n\omega)$.
 In terms of the total dipole moment ${\bf p}\equiv(r_>-r_<)(z_>-z_<)[2\pi c/(n\omega)]{\bf P}$
 of the contributing source, therefore, Eq.\ (74) can be written as
$$\vert{\tilde{\bf E}}_n\vert\sim\textstyle{1\over2}{r_P}^{-1}(n\omega/c)^2\vert n\, J_{n-\Omega/\omega}(n){\bf p}\vert.\eqno(75)$$ 
This differs from the familiar expression for the radiation field of a stationary
dipole in the plane normal to its direction~\cite{AAS11}
by the factor $\vert{1\over2} n\, J_{n-\Omega/\omega}(n)\vert$,
a factor which is of the order of $n^{2\over3}$ when $n\gg1$.
The difference can clearly be traced to the phasing of the array of
oscillating dipoles that constitute the present moving source (cf.\ Appendix A):
the field calculated in Section~\ref{V} receives contibutions from an interval of retarded
time that exceeds the period $2\pi/(n\omega)$ of the oscillations of its source by the very same factor.
\section{Efficiency of the radiative process}
\label{VIII}
The efficiency of the emission process analyzed in Sections~\ref{V} and \ref{VI}
is essentially independent of the way in which a polarization current with a superluminally
rotating distribution pattern is created.
Our purpose in this section is to derive a general expression for
estimating the radiation efficiency in the high-frequency regime $n\sim(\Omega/\omega)^3$ and
to apply the resulting expression to the particular method of
implementing the source density (7) that is described in Appendix A.

If the polarization current density ${\bf j}$ that acts as the source of the present
radiation is produced by the influence of an external electric field ${\bf E}_{\rm ext}$
on a polarizable medium with electric susceptibility $\chi_e$, then the power
required for maintaining ${\bf j}$ within a volume $V$ would be $P_{\rm in}=\int_V{\bf j\cdot E}_{\rm ext}d^3x$.
The induced polarization ${\bf P}$ is given by $\chi_e{\bf E}_{\rm ext}$, so
that the polarization current ${\bf j}=\partial{\bf P}/\partial t$ would have the
magnitude $\vert{\bf j}\vert\sim \chi_e\Omega\vert{\bf E}_{\rm ext}\vert$,
where $\Omega$ is the dominant frequency in the spectrum of oscillations
of ${\bf E}_{\rm ext}$ and hence ${\bf j}$.
The input power would therefore be of the order of
$$P_{\rm in}\sim L_rL_{\hat\varphi}L_z \vert{\bf j}\vert^2/(\chi_e\Omega)\eqno(76)$$ 
in terms of $\vert{\bf j}\vert$, where we have expressed $V$ as the product
$L_rL_{\hat\varphi}L_z$ of the length scales of the source distribution in various directions.

The power that is emitted into the frequency band $(\Delta n)\omega$ centred
at $n\omega\sim(\Omega/\omega)^2\Omega$ is, according to the analysis in Section~\ref{V}, given by
$$P_{\rm out}\sim c^{-1}{L_r}^2{L_z}^2\vert{\bf j}\vert^2n^{-2}\Delta n\Delta\Omega_P,\eqno(77)$$
where the solid angle $\Delta\Omega_P$ is an estimate of the size of the beam that is emitted
into the plane of source's orbit.  Note that $r_>-r_<$, $z_>-z_<$ and $\Omega\vert{\bf s}\vert$ in Eq.\ (57)
correspond to $L_r$ and $L_z$, and $\vert{\bf j}\vert$, respectively,
and that the high-frequency limit of $J_n(n)$ has the same order of magnitude as $(\Omega/\omega)^{-1}\sim n^{-{1\over3}}$.

The above two expressions for the input and output powers imply that the efficiency of the emission process in question has a value of the order of
$$P_{\rm out}/P_{\rm in}\sim \chi_e(L_r/L_{\hat\varphi})(L_z\omega/c)n^{-{5\over3}}\,\Delta n\,\Delta\Omega_P,\eqno(78)$$
in which we have replaced $\Omega/\omega$ by $n^{1\over3}$.
The interval $\Delta n$ over which the high-frequency component of the radiation is emitted is of the same
order of magnitude as $n$ [cf.\ Figs.\ 6--9].
But the solid angle $\Delta\Omega_P$ in this expression is only a small fraction of $4\pi$ (see below):  the estimate in
Eq.\ (77) is valid only within the distance $L_z$ of the plane of the source's orbit.

Radiation of frequency $n\omega\sim(\Omega/\omega)^2\Omega$ can be detected also outside
the plane of source's orbit.  Although the value of $\Delta\Omega_P$ in the corresponding
expression for $P_{\rm out}/P_{\rm in}$ is of the order of unity when $\theta_P\ne\pi/2$ (see Section~\ref{VI}),
the greater steepness of the spectrum reduces the efficiency in this case:
the dependence of $P_{\rm out}/P_{\rm in}$ on $n$ is by a factor of the order of
$\vert{\bf Q}_z\vert^2\sim n^{-2}$ smaller than that found in Eq.\ (78) [see Eq.\ (69)].

Consider an experimental device, such as that described in Appendix A, that is built with a polarizable medium
with the electric susceptibility $\chi_e\simeq 10$ and electrodes that have the dimensions
$L_r\simeq L_{\hat\varphi}\simeq 5$ cm, $L_z\simeq 1$~cm.
If this device is operated with $\omega/(2\pi)\simeq5$ MHz and $\Omega/(2\pi)\simeq300$~MHz
(the exact value of $\Omega/\omega$ being different from an integer),
then the efficiency with which the radiation of frequency $n\omega/(2\pi)\sim(\Omega/\omega)^2\Omega/(2\pi)\simeq1$~THz
is generated by each electrode would be of the order of $P_{\rm out}/P_{\rm in}\simeq3\times10^{-6}\Delta\Omega_P$.
Note that $(n/\pi)(L_z\omega/c)^2\simeq10^{-1}$ and $(n/\pi)(L_r\omega/c)^2\simeq1$
in this case, so that the boundary between the Fresnel and far zones
lies at a distance from the source that is shorter than the radius $c/\omega\simeq10^3$ cm of the light
cylinder [see Eq.\ (55)].  Beyond the Fresnel zone, the radiation that is beamed
into the cylindrical region $z_<\le z_P\le z_>$ surrounding the plane of source's
orbit therefore subtends a solid angle that is smaller than $\Delta\Omega_P\sim L_z\omega/c\sim10^{-3}$ radians.
\section{Concluding remarks}
\label{IX}
We have investigated the spectral features of the intense localized electromagnetic waves that
are generated by volume polarization currents with superluminally-moving distribution
patterns. The analysis is based upon current practical devices for
investigating emission from accelerated superluminal sources~\cite{AAS12,johan,test}; these devices (Appendix~A) produce polarization currents whose
distribution patterns rotate and oscillate with two incommensurate frequencies ($\omega$ and $\Omega$).  Although the only frequencies entering the production of the emitting currents are $m\omega$ and $\Omega$ (see Table~1), we find that the broadband signals from such devices contain frequencies that are higher than the oscillation frequency $\Omega$ by a factor of the order of $(\Omega/\omega)^2$.  This does not mean that the linear emission process considered here is capable of generating an output at frequencies that are not carried by its input (the source), thus violating the convolution theorem.  What is made possible by this process is to generate radiation of a certain frequency from a source whose {\it creation} does not require that frequency.  The {\it spectra} of such sources do contain the emitted frequencies [see Eq.\ (9)].
 
The high frequencies (none of which are required for the practical implementation of the source)
stem from the cooperation of the following two effects.  The retarded time is a multi-valued function
of the observation time in the superluminal regime, so that the interval of
retarded time during which a particular set of wave fronts is emitted by a
a volume element of the source can be significantly longer than the interval of observation time during
which the same set of wave fronts is received at the observation point.
In addition, a remarkable effect of centripetal acceleration is to enrich the spectral content
of a rotating volume source, for which $\Omega/\omega$ is different from an integer,
by effectively endowing the distribution of its density with space-time discontinuities.
These results are mathematically rigorous consequences of the familiar classical
expression for the retarded potential.

The spectral distribution for the emitted radiation is summarised in Figs.\ 8 and 9,
and its possible polarization states are listed in Table~2.

Many features of the radiation that is discussed in the present paper are shared by the high-frequency
emissions that would arise from superluminally moving sources with generically different trajectories.
An example is a superluminal source which moves along a straight line with acceleration.
In the case of the circularly-moving superluminal source considered here,
the parameter $\omega$ determines both the linear velocity of each source element $(r\omega)$
and its acceleration $(r\omega^2)$.  For a superluminal source whose distribution pattern
moves rectilinearly, on the other hand, velocity ($v$) and acceleration ($a$) are two independent parameters.

The emission ($\delta t$) and reception ($\delta t_P$) time intervals for the waves arising from the
source elements that approach the observer with the wave speed and zero
acceleration are in the rectilinear case, too, related by a cubic equation:
$$\delta t_P=\textstyle{1\over2}(a/c)^2[(v/c)^2-1]^{-1}(\delta t)^3+\cdots,\eqno(79)$$
in which $v$ is the retarded value of the source velocity (see Appendix D of~\cite{AAS1}).
Once again, therefore, a moving source whose strength fluctuates like $\cos(\Omega t)$
would generate a field which oscillates with a period different from that of its source:
with the period $[2\pi a/(c\Omega)]^2[(v/c)^2-1]^{-1}(2\pi/\Omega)$.

In practice, however, there is a crucial difference between rectilinear and centripetal accelerations.
Centripetal acceleration, as we have seen, enriches the spectral content of a rotating volume source at the same time
as giving rise to the formation of caustics and so the compression of $\delta t_P$ relative to $\delta t$.
In contrast, rectilinear acceleration requires for its implementation the very frequencies
it endows the source with:  the range of frequencies with which the amplitude of a linearly
accelerated source oscillates at a fixed point on its path (within its distribution)
is as wide as the range of frequencies that feature in the Fourier decomposition of its density with respect to time.

The rotating superluminal source described in Eq.\ (7) can be implemented by an experimentally-viable
device~\cite{johan,test} whose construction and operation only entail oscillations at the two frequencies
$m\omega$ and $\Omega$ (see Appendix A).  As a source of radiation at
frequencies which cannot be normally generated in the laboratory,
except by means of large-scale facilities such as synchrotrons or free electron lasers,
the potential practical significance of such a device is clearly enormous~\cite{AAS12};
as we have shown in Section~\ref{VIII},
the efficiency is large enough for many spectroscopic applications to be viable.
\section{Acknowledgements}
The authors acknowledge support from EPSRC Research Grant No.\ GR/M52205, and  H.\ A.\ thanks J.\ M.\ Rodenburg and
D.\ Lynden-Bell for stimulating and helpful discussions. W. Hayes and G.S. Boebinger are thanked for their
encouragement.
\section{Appendix A: practical implementation of the source}
The purpose of this appendix is to demonstrate that the polarization described in Eq.\ (7)
can be implemented by an experimentally viable device~\cite{johan,test} whose construction and operation only entail oscillations at the two frequencies $m\omega$ and $\Omega$.

Consider a circular ring of radius $r$, made of a dielectric material, with an array of $N$ electrode pairs that are placed beside
each other around its circumference (Fig.\ 10).  With a sufficiently large value of $N$ (to be determined below), it
would be possible to generate a sinusoidal distribution of polarization along the length of the
dielectric by applying a voltage to each pair independently.
The distribution pattern of this polarization can then be animated, i.e.\ set in motion,
by energizing the electrodes with time-varying signals.
We can synthesize a transverse polarization wave $\cos[m(\varphi-\omega t)]$ moving around the
ring by driving each electrode pair with a sinusoidal signal
whose frequency is fixed but whose phase depends on the position of the pair around the ring.
\begin{figure}[tbp]
   \centering
\includegraphics[height=7cm]{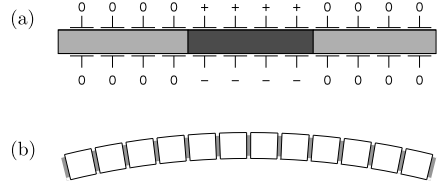}
\caption{View of the experimental device (a) from the side and
(b) from above showing an arc of the dielectric medium (dotted region), its
polarized part (darker dotted region), and the electrode pairs
(designated by `$\pm$' where on and by `$00$' where off).}
\end{figure}

The frequency $m\omega$ and the wavelength $2\pi r/m$ of the travelling polarization wave,
and hence its speed $r\omega$, can be controlled at will by varying the frequency $m\omega$
with which the electrodes are driven and the phase difference $2\pi m/N$ between neighboring electrode pairs.
Introducing the factor $\cos(\Omega t)$ of Eq.\ (7) simply corresponds to mixing a second
frequency into the signal driving the electrodes with a phase that is the same for all electrodes.

To estimate the required value of $N$, let us note that the $(\varphi,t)$ dependence
of the polarization that is thus generated by the discrete set of electrodes described above has the form
$$P(\varphi,t)=\cos(\Omega t)\sum_{k=0}^{N-1}\Pi(k-N\varphi/2\pi)\cos[m(\omega t-2\pi k/N)],\eqno(A1)$$
in which $\Pi(x)$ denotes the rectangle function,
a function that is $1$ when $\vert x\vert<\textstyle{1\over2}$ and
zero when $\vert x\vert>\textstyle{1\over2}$.  [For any given $k$,
the function $\Pi(k-N\varphi/2\pi)$ is non-zero only over the interval $(2k-1)\pi/N<\varphi<(2k+1)\pi/N$.]
When the electrodes operate over a time interval exceeding $2\pi/\omega$,
the generated current is a periodic function of $\varphi$ for
which the range of values of $\varphi$ correspondingly exceeds the period $2\pi$.

The Fourier series representation of the function $\Pi(k-N\varphi/2\pi)$ with the period $2\pi$ is given by
$$\Pi(k-N\varphi/2\pi)=N^{-1}+\sum_{n=1}^\infty (n\pi/2)^{-1}\sin(n\pi/N)\cos[n(\varphi-2\pi k/N)].\eqno(A2)$$
If we now insert Eq.\ (A2) in Eq.\ (A1) and use formula (4.3.32) of~\cite{AAS9}
to rewrite the product of the two cosines in the resulting expression as the sum of two cosines, we obtain two
infinite series each involving a single cosine and extending over $n=1, 2, \cdots, \infty$.
These two infinite series can then be combined (by replacing $n$ in one of them with $-n$
everywhere and performing the summation over $n=-1, -2, \cdots, -\infty$) to arrive at
%$$\eqalignno{
\begin{eqnarray*}
P(\varphi,t)=&\cos(\Omega t)\sum_{n=-\infty}^{\infty}(n\pi)^{-1}\sin(n\pi/N)\qquad\qquad\qquad\qquad\cr
&\times\sum_{k=0}^{N-1}\cos[m\omega t-n\varphi+2\pi(n-m)k/N],&(A3)
\end{eqnarray*}
%\cr}$$
in which the order of summations with respect to $n$ and $k$ have been interchanged
and the contribution $N^{-1}$ on the right-hand side of Eq.\ (A2) has been incorporated
into the $n=0$ term:  the coefficient $(n\pi)^{-1}\sin(n\pi/N)$ has the value $N^{-1}$ when $n=0$.

The finite sum over $k$ can be evaluated by means of the geometric progression.  The result, according to formula (1.341.3) of \cite{AAS13}, is
%$$\eqalignno{
\begin{eqnarray*}
\sum_{k=0}^{N-1}\cos[m\omega t-n\varphi+2\pi(n-m)k/N]=&\cos[m\omega t-n\varphi+\pi(n-m)(N-1)/N]\qquad\qquad\cr
&\times\sin[(n-m)\pi]\csc[(n-m)\pi/N].\qquad\qquad&(A4)
\end{eqnarray*}
%\cr}$$
The right-hand side of Eq.\ (A4) vanishes when $(n-m)/N$ is different from an integer.  If $n=m+lN$, where $l$
is an integer, the above sum would have the value $N\cos(m\omega t-n\varphi)$, as can be
seen by directly inserting $n=m+lN$ in the left-hand side of Eq.\ (A4).
Performing the summation with respect to $k$ in Eq.\ (A3), we therefore obtain
%$$\eqalignno{
\begin{eqnarray*}
P(\varphi,t)=&(m\pi/N)^{-1}\sin(m\pi/N)\cos(\Omega t)\big\{\cos[m(\varphi-\omega t)]\qquad\qquad\qquad\qquad\cr
&+\sum_{l\neq 0}(-1)^l(1+Nl/m)^{-1}\cos[(Nl+m)\varphi-m\omega t]\big\},\qquad\qquad&(A5)
\end{eqnarray*}
%\cr}$$
since only those terms of the infinite series survive for which $n$ has the value $m+lN$ with
an $l$ that ranges over all integers from $-\infty$ to $\infty$.

We have written out the $l=0$ term of the series in Eq.\ (A5) explicitly in order
to bring out the following points.  The parameter $N/m$, which signifies the number
of electrodes within a wavelength of the source distribution, need not be large for
the factor $(m\pi/N)^{-1}\sin(m\pi/N)$ to be close to unity:  this factor
equals $0.9$ even when $N/m$ is only $4$.
Moreover, if the travelling polarization wave $\cos(m{\hat\varphi})$ that is associated
with the $l=0$ term has a phase speed $r\omega$ that is only moderately superluminal,
the phase speeds $r\omega/\vert1+Nl/m\vert$ of the waves described by all the other terms in
the series would be subluminal.  Not only would these other polarization waves have amplitudes
that are by the factor $\vert1+Nl/m\vert^{-1}$ smaller than that of the fundamental wave
associated with $l=0$, but they would generate electromagnetic fields whose characteristics
(e.g.\ the peak frequencies of their spectra) are different from those generated by the superluminally moving polarization wave.

The fundamental ($l=0$) component of the polarization current that is created by the
present device, therefore, has precisely the same $(\varphi, t)$ dependence as that which is described in
Eq.\ (7) of the main text.  Neither the reduction in its amplitude, that arises from
the departure of the value of $(m\pi/N)^{-1}\sin(m\pi/N)$ from unity, nor the presence of the
other lower amplitude waves that are superposed on it make any difference
to the fact that the creation of this fundamental component only
entails the two frequencies $m\omega$ and $\Omega$.
Linearity of the emission process ensures that the radiation generated by an individual term of
the series in Eq.\ (A5) is not in any way affected by those that are generated by the other terms of this series.

For the distribution pattern of the created polarization current to be moving, it is however essential that the number of electrodes per
wavelength of this pattern, $N/m$, should exceed 2.  For $N/m=2$, the $l=-1$
term becomes $\cos[m(\varphi+\omega t)]$ and so represents a wave that has the same amplitude as,
and travels with the same speed in the opposite direction to, the wave represented by the $l=0$ term.
In this particular case, the fundamental wave is thus turned into a standing wave.

Note, finally, that the speed of light is easily attainable:  if $N=100$ electrodes on a circle of radius
$r=1$ m are driven with the frequency $m\omega=600$ MHz and the phase difference
$2\pi m/N=0.125$ radians, then the distribution pattern of the induced polarization current would consist of
$m=2$ wavelengths of a sinusoidal wave train which would move along the
circle with the speed $r\omega=3\times10^8$ m/s.  Moreover, only an arc of such a circularly
shaped device is needed for generating the present radiation:  the pulse that is received at any given
observation point arises almost exclusively from the limited part of the source which
approaches the observer with the speed of light and zero acceleration at the retarded time.

\end{document}